\documentclass[journal]{IEEEtran}
\usepackage{multirow}
\usepackage{color}
\usepackage[table]{xcolor}
\usepackage[font=small]{caption}
\usepackage{subfloat}
\usepackage{verbatim}
\usepackage{csquotes}
\usepackage{cite}
\ifCLASSINFOpdf
   \usepackage[pdftex]{graphicx}
\else
   \usepackage[dvips]{graphicx}
\fi

\usepackage{mathtools}

\usepackage{tabularx,booktabs}
\newcolumntype{Y}{>{\centering\arraybackslash}X}
\newcolumntype{Z}{>{\small\centering\arraybackslash}X}
\newcolumntype{S}{>{\hsize=.9\hsize}X}
\newcolumntype{B}{>{\hsize=1.1\hsize}X}

\usepackage[outdir=./]{epstopdf}
\usepackage{amsfonts}
\usepackage{amsmath}
\usepackage{pgfplots}
\usepackage{algorithm,algpseudocode}
\usepackage{array}
\usepackage{graphicx}
\usepackage{subcaption}
\usepackage{fixltx2e}
\usepackage{bbm}
\usepackage{stfloats}
\usepackage{algorithm}
\usepackage{url}
\usepackage{upgreek}
\usepackage{amsthm}
\usepackage{balance}
\usepackage{enumitem}

\providecommand{\parab}[1]{\noindent\textbf{#1}}
\definecolor{myb}{rgb}{0, 0, 0}

\providecommand{\parab}[1]{\noindent\textbf{#1}}
\providecommand{\para}[1]{\noindent\textit{#1}}

\providecommand{\parab}[1]{\noindent\textbf{#1}}
\providecommand{\para}[1]{\noindent\textit{#1}}

\newtheorem{theorem}{Theorem}
\newtheorem{lemma}{Lemma}

\allowdisplaybreaks

\begin{document}

\title{Network-Aware Optimization of \\ Distributed Learning for Fog Computing}

\author{Su Wang, Yichen Ruan, Yuwei Tu, Satyavrat Wagle, Christopher G. Brinton, and Carlee Joe-Wong

\thanks{This work was presented in part at the 2020 IEEE Conference on Computer Communications (INFOCOM) \cite{tu2020network}.}
\thanks{S. Wang and C. Brinton are with the School of Electrical and Computer Engineering at Purdue University. email: \{wang2506,cgb\}@purdue.edu}
\thanks{Y. Ruan, S. Wagle, and C. Joe-Wong are with the Department of Electrical and Computer Engineering at Carnegie Mellon University. email: \{yichenr, srwagle, cjoewong\}@andrew.cmu.edu}}
\maketitle

\begin{abstract}
Fog computing promises to enable machine learning tasks to scale to large amounts of data by distributing processing across connected devices. Two key challenges to achieving this goal are (i) heterogeneity in devices' compute resources and (ii) topology constraints on which devices communicate with each other. We address these challenges by developing a novel network-aware distributed learning methodology where devices optimally share local data processing and send their learnt parameters to a server for periodic aggregation. Unlike traditional federated learning, our method enables devices to offload their data processing tasks to each other, with these decisions optimized to trade off costs associated with data processing, offloading, and discarding. We analytically characterize the optimal data transfer solution under different assumptions on the fog network scenario, showing for example that the value of offloading is approximately linear in the range of computing costs in the network when the cost of discarding is modeled as decreasing linearly in the amount of data processed at each node. Our experiments on real-world data traces from our testbed confirm that our algorithms improve network resource utilization substantially without sacrificing the accuracy of the learned model, for varying distributions of data across devices. We also investigate the effect of network dynamics on model learning and resource costs.
\end{abstract}


\IEEEpeerreviewmaketitle

\section{Introduction}
\label{sec:intro}

New technologies like autonomous cars and smart factories are coming to rely extensively on data-driven machine learning (ML) algorithms~\cite{cisco-5giiot,chatzopoulos2017mobile,ieee-5g-health} to produce near real-time insights based on historical data. Training ML models at realistic scales, however, is challenging, given the enormous computing power required to process today's data volumes. The collected data is also dispersed across networks of devices, while ML models are traditionally managed in a centralized manner \cite{wang2019adaptive}.

Fortunately, the rise in data generation in networks has occurred alongside a rise in the computing power of networked devices.
Thus, a possible solution for real-time training of and inferencing from data-driven ML algorithms lies in the emerging paradigm of fog computing, which aims to design systems, architectures, and algorithms that leverage device capabilities between the network edge and cloud \cite{chiang2016fog}. 
Deployment of 5G wireless networks and the Internet of Things (IoT) is accelerating adoption of this computing paradigm by expanding the set of connected devices with compute capabilities and enabling direct device-to-device communications between them \cite{IEEE-D2D}. Though centralized ML algorithms are not optimized for such environments, distributed data analytics is expected to be a major driver of 5G adoption \cite{ieee-5g-analytics}.

\begin{figure}
\vspace{-0.15in}
\centering
\begin{subfigure}{0.16\textwidth}
\includegraphics[width = \textwidth]{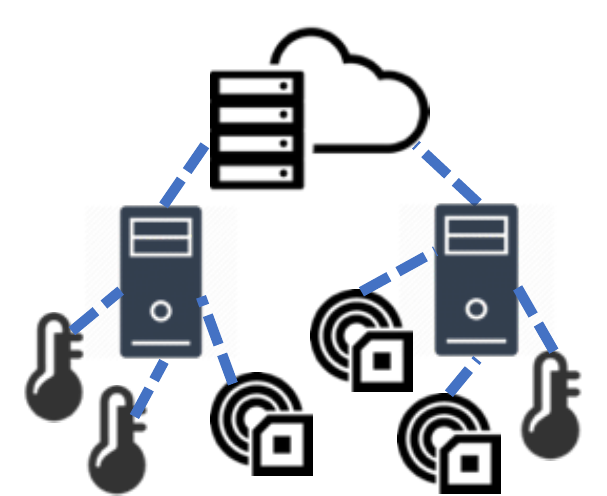}
\caption{Hierarchical}
\label{fig:hierarchy}
\end{subfigure}
\hspace{0.05\textwidth}
\begin{subfigure}{0.16\textwidth}
\includegraphics[width = \textwidth]{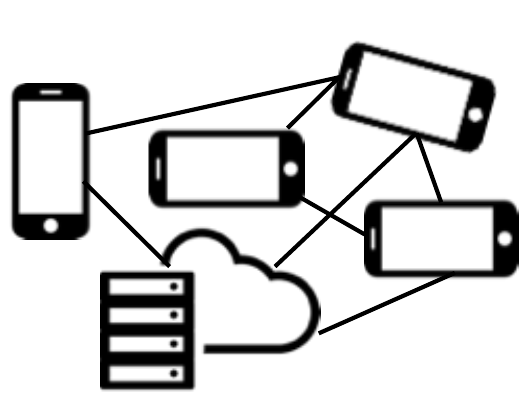}
\caption{Social network}
\label{fig:social}
\end{subfigure}
\caption{Cartoon illustrations of two example topologies for fog computing that we consider. In the hierarchical case, less powerful devices are connected to more powerful ones, while for the social network, connections are denser and devices tend to be similar.}
\label{fig:topologies}
\vspace{-0.15in}
\end{figure}

Initial efforts in decentralizing ML have focused on decomposing model parameter updates over several nodes, typically managed by some centralized server \cite{pu2018push,mcmahan2017communication}. Most of these methods implicitly assume idealized network topologies where node and link properties are homogeneous. Fog environments, by contrast, are characterized by devices' computation and communication resource heterogeneity, e.g., due to power constraints or privacy considerations. 
For example, consider the two common fog topologies depicted in Figure~\ref{fig:topologies}. In the hierarchical scenario, weaker edge devices are connected to powerful edge servers. In the social network case, devices tend to have similar compute resources, but connectivity may vary significantly depending on levels of trust between users \cite{yang2018learning}. 

A central question that arises, then, in adapting ML methodologies to fog environments is: \emph{How should each fog device contribute to the ML training and inference?} The examples above motivate techniques for \emph{network-aware} distributed learning, which (i) account for the potentially heterogeneous computation and communication resources across devices, and (ii) leverage the network topology for device communications to optimize the distribution of data processing through the network. In this paper, we develop a novel network-aware distributed learning methodology that optimizes the distribution of processing across a network of fog devices.

\vspace{-0.1in}
\subsection{Machine Learning in Fog Environments}
\label{subsec:intro:scenarios}
ML models are generally trained by iteratively updating estimates of parameter values, such as weights in a neural network, that best ``fit'' empirical data, through data processing at one or more devices. We face two major challenges in adapting such training to fog networking environments: \emph{(i) heterogeneity in devices' compute resources} and \emph{(ii) constraints on devices' abilities to communicate with each other}. We outline these characteristics, and the potential benefits enabled by our network-aware distributed learning methodology, in some key applications below:

\parab{Privacy-sensitive applications.} Many ML applications learn models on sensitive user data, e.g., for health monitoring \cite{ieee-5g-health}. Due to privacy concerns, most of these applications have devices train their models on local data to avoid revealing data to untrustworthy nodes \cite{yang2018learning}. Offloading ML data processing to trusted devices, e.g. if one user owns multiple devices, can reduce training times and improve model accuracy. 

\parab{Internet-connected vehicles} can collaboratively learn about their environment~\cite{cisco-5giiot}, e.g., by combining their data with that of road sensors to infer current road or traffic conditions. Since sensors have less computing capabilities than vehicles, they will likely offload their data to vehicles or roadside units for processing. This offloading must adapt as vehicles move and their connectivity with (stationary) road sensors changes. 

\parab{Augmented reality (AR)} uses ML algorithms for e.g., image recognition \cite{chatzopoulos2017mobile} to overlay digital content onto users' views of an environment. A network of AR-enabled devices can distributedly train ML models, but may exhibit significant heterogeneity: they can range from generic smartphones to AR-specific headsets, with different battery levels. As the AR users move, connectivity between devices will also change.

\parab{Industrial IoT.} 5G networks will allow sensors that power control loops in factory production lines to communicate across the factory floor~\cite{cisco-5giiot,ashraf2016ultra}, enabling distributed ML algorithms to use this data for, e.g., predicting production delays. Determining which controllers should process data at specific sensors is an open-question: it depends on sensor-controller connectivities, which may vary with factory activity. 

\vspace{-0.1in}
\subsection{Outline and Summary of Contributions}
First, Section \ref{sec:related} differentiates our work from relevant literature. To the best of our knowledge, we are the first to optimize the distribution of ML data processing (i.e., training) tasks across fog nodes, leading to several contributions:

\parab{Formulating the task distribution problem} (Section~\ref{sec:formulation}). In deciding which devices should process which datapoints, our formulation accounts for resource limitations and model accuracy. While ideally more of the data would be processed at devices with more computing resources, sending data samples to such devices may overburden the network. Moreover, processing too many data samples can incur large processing costs relative to the gain in model accuracy. 
We derive new bounds (Theorem~\ref{thm:error}) on the model accuracy when data can be moved between devices, showing that the optimal task distribution problem can be formulated as a convex optimization.

\parab{Characterizing the optimal task distribution} (Section~\ref{sec:analysis}). Solving the optimization in Sec.~\ref{sec:formulation} requires specifying network characteristics that may be unknown in advance. We analyze the expected deviations from our assumptions in Sec.~\ref{sec:formulation} to derive guidelines on estimating these characteristics (Theorem~\ref{thm:compute-exp}). We then consider two different models of discard cost, one linear and one convex in the number of datapoints processed, and use them to derive the optimal distribution of data processing through the network for typical fog topologies (Theorems~\ref{thm:unconstrained} and~\ref{thm:nonlinear}). Further, we use the result from the linear case to estimate the reduction in processing costs due to data movement (Theorem~\ref{thm:cost_uniform}). Overall, these results are the first to characterize the integration of offloading into federated learning over graph topologies.

\parab{Experimental validation} (Section~\ref{sec:implementation}). We train classification models on the MNIST dataset to validate our algorithms. We use a Raspberry Pi testbed to emulate network delays and available compute resources, under both i.i.d. and non-i.i.d. device data. Our proposed algorithm nearly halves the processing overhead yet achieves an accuracy comparable to centralized model training, which validates the benefits of network-aware distributed learning. The experiments also reveal how the network connectivity, size, and topology structure impact the trained model accuracy and network resource costs.

\section{Related Work}
\label{sec:related}

We contextualize our work within prior results on (i) federated and distributed machine learning, and (ii) methods for offloading ML tasks from mobile devices to edge servers.

\vspace{-0.1in}
\subsection{Distributed and Federated Learning}
In classical distributed learning, ``workers'' each compute a parameter value from their local data. These results are aggregated at a central server, and updated parameter values are sent to the workers for the next round of local computations. Several works have considered how network topologies can affect ML training. 
Specifically,~\cite{he2018cola} optimizes decentralized training of linear models when workers exchange parameters with their neighbors instead of a central server, while \cite{predd2009collaborative} studied D2D (device-to-device) message passing that emulates global communications with a server, and~\cite{Tang2019DistributedLO} optimizes performance when communications can fail. 
In fog networks, devices may not have the resources to send updates to a server at every time period~\cite{wang2019adaptive}, so we focus on the newer federated learning methodology for distributed model training.

Devices in federated learning perform a \emph{series} of local updates between aggregations, and send their resulting model updates to the server~\cite{dutta2018slow,konevcny2016federated,mcmahan2017communication}. This framework preserves user privacy by keeping data at local devices~\cite{shokri2015privacy} and reduces the amount of communication between devices and the central server. Since its proposition in \cite{mcmahan2017communication}, federated learning has generated significant research interest; see \cite{rahman2020survey} for a comprehensive survey. Compared to traditional distributed learning, federated learning introduces two new challenges: firstly, data may not be identically and independently distributed as devices locally generate the samples they process; and secondly, in fog/edge networks, devices may have limited ability to perform and communicate local updates due to resource constraints. Many works have attempted to address the first challenge; for instance, \cite{smith2017federated} trains user-specific models within a multi-task learning framework, while \cite{zhao2018federated} showed that sharing small subsets of user data can significantly increase the accuracy of a single model. Recent efforts have also considered optimizing the frequency of parameter aggregations according to a fixed budget of network and computing resources~\cite{wang2019adaptive}, or adopting a peer-to-peer decentralized learning framework~\cite{neglia2019role}. 

Our work is more related to the second challenge of resource heterogeneity. Existing works have focused on minimizing the communication costs between the workers and the server. Specifically, \cite{konen2016federated} proposes to reduce uplink costs by restricting and compressing the parameter space prior to transmission. \cite{strom2015scalable} proposes a method for thresholding updates for transmission, while the method in \cite{Aji_2017} only communicates the most important individual gradient results, which \cite{sattler2019robust} extends to compress both downlink and uplink communication. To reduce the number of uplink transfers,~\cite{chen2019communicationefficient,lalitha2019peertopeer} aggregate subsets of the network. For wireless networks in particular, \cite{tran2019federated} proposes methods to minimize power consumption and training time.

Unlike these works, we focus on the network topology between nodes to optimize the tradeoffs among communication, computation, and model performance in federated learning. Specifically, our work leverages D2D communications for offloading data from resource-constrained to resource-rich nodes, as is present in new 5G and IoT network technologies~\cite{chiang2016fog}. In integrating D2D offloading with federated learning, we derive novel results on optimizing the distribution of data processing over a fog computing system to train ML models.

\vspace{-0.1in}
\subsection{Offloading} Fog computing introduces opportunities to pool network resources and maximize the use of idle computing/storage in completing resource-intensive activities \cite{chiang2016fog}. Offloading mechanisms improve system performance when high-bandwidth network connections are available between devices. Offloading has significantly accelerated ML tasks such as linear regression training \cite{chang2017decomposing} and neural network inference \cite{ran2018deepdecision}. Existing literature has also considered splitting different layers in deep neural networks between fog devices and edge/cloud servers. Specifically, \cite{hu2019dynamic} proposed two network-dependent schemes and a min-cut problem to accelerate the training phase, while \cite{teerapittayanon2017distributed} developed an architecture to intelligently locate models on local devices or the cloud based on network reliability.

Prior works on deep learning offloading in fog/edge computing generally focus on network factors, such as cost/latency, rather than model accuracy. For example, \cite{xu2019heuristic,li2018learning} maximize throughput in wireless and sensor IoT networks. \cite{sun2020energy} studies D2D data offloading in federated learning, but focuses on communication protocols to facilitate over-the-air parameter aggregations and dataset duplication across nodes. Our methodology considers more general ML models, optimizes tradeoffs between network cost and model accuracy, and provides novel theoretical performance bounds.

\section{Model and Optimization Formulation}
\label{sec:formulation}

We first define our model for fog networks (Section~\ref{subsec:formulation:prelim}) and machine learning training (Section~\ref{subsec:formulation:learn}), and then formulate the task distribution optimization problem (Section~\ref{subsec:formulation:opt}).

\vspace{-0.1in}
\subsection{Fog Computing System Model}\label{subsec:formulation:prelim}
We consider a set $V$ of $n$ fog devices forming a network, an aggregation server $s$, and discrete time intervals $t = 1,\ldots,T$ as the period for training an ML model. Each device, e.g., a sensor or smartphone, can both collect data and process it to contribute to the ML task. The server $s$ aggregates the results of each device's local analysis, as will be explained in Section \ref{subsec:formulation:learn}. Both the length and number of time intervals may depend on the specific ML application. In each interval $t$, we suppose a subset of devices $V(t)$, indexed by $i$, is active (i.e., available to collect and/or process data). For simplicity of notation, we omit $i$'s dependence on $t$. 

\subsubsection{Data collection and processing} We use $D_i(t)$ to denote the set of data collected by device $i \in V(t)$ for the ML task at time $t$; $d \in D_i(t)$ denotes each datapoint. Note, $D_i(t) = \emptyset$ if a device does not collect data at time $t$. $G_i(t)$, by contrast, denotes the set of datapoints \emph{processed} by each device at time $t$, for the ML task. 
In conventional distributed learning frameworks, $D_i(t) = G_i(t)$, as all devices process the data they collect~\cite{wang2019adaptive}; separating these variables is one of our main contributions. We suppose that each device $i$ can process up to $C_i(t)$ datapoints at each time $t$, incurring a cost of $c_i(t)$ for each point. 
For example, devices with low battery will have lower capacities $C_i(t)$ and higher costs $c_i(t)$.

\subsubsection{Fog network connectivity} The devices $V$ are connected to each other via a set $E$ of directed links $(i,j)$ between devices $i$ and $j$; $E(t) \subseteq E$ denotes the set of functioning links at time $t$. The overall system can then be described as a directed graph $(\left\{s,V\right\},E)$ with vertices $V$ representing the devices and edges $E$ the links between them. We suppose that $(\left\{s,V(t)\right\}, E(t))$ is connected at each time $t$ and that links between devices are single-hop, i.e., devices do not use each other as relays except possibly to communicate with the server. 
The scenarios outlined in Section \ref{subsec:intro:scenarios} each possess such an architecture: in smart factories, for example, a subset of the floor sensors connect to each controller. 
Each link $(i,j)\in E(t)$ is characterized by a capacity $C_{ij}(t)$, i.e., the maximum datapoints transferable in a time interval, and a per-unit ``cost of connectivity'' $c_{ij}(t)$. This cost may reflect network conditions or a desire for privacy, and will be high if sending from $i$ to $j$ is less desirable at $t$.  

\subsubsection{Data structure} Each datapoint $d$ takes the form $(x_d, y_d)$, where $x_d$ is an attribute/feature vector and $y_d$ is an associated label.\footnote{For unsupervised ML tasks, there are no labels $y_d$. However, we can still minimize a loss defined in terms of the $x_d$, similar to (1).} We use $D_V = \cup_{i,t} D_i(t)$ to denote the full set of datapoints collected by all devices over all time. Following prior work~\cite{yang2013trading,zhang2013information}, we model data collection as each device $i$ sampling points uniformly at random from a (usually unknown) distribution $\mathcal{D}_i$. Temporal changes in $\mathcal{D}_i$ are assumed to be slow compared to the time horizon $T$. We use $\mathcal{D} = \cup_i \mathcal{D}_i$ to denote the global distribution induced by these $\mathcal{D}_i$. Our model implies that the relationship between $x_d$ and $y_d$ is temporally invariant, which is common in the applications discussed in Section~\ref{subsec:intro:scenarios}, e.g., image recognition from road cameras at fixed locations or AR users with random mobility patterns. We use such a dataset for evaluation in Section \ref{sec:implementation}.

\vspace{-0.1in}
\subsection{Machine Learning Model}\label{subsec:formulation:learn}
Our goal is to learn a parameterized model that outputs $y_d$ given the input feature vector $x_d$. We use the vector $w$ to denote the set of model parameters, whose values are chosen so as to minimize a loss function $L(w | \mathcal{D})$ that is defined for the specific ML model (e.g., squared error for linear regression, cross-entropy loss for multi-class classifiers \cite{murphy2012machine}). Since the overall distributions $\mathcal{D}_i$ are unknown, instead of minimizing $L(w | \mathcal{D})$ we minimize the empirical loss function, as commonly done in ML model training (e.g.,~\cite{wang2019adaptive},~\cite{mcmahan2017communication}):
\begin{equation}
\label{eq:glob}
\underset{w}{\text{minimize}} \;\; L(w | D_V) = \frac{\sum_{t = 1}^T \sum_{i \in V(t)} \sum_{d \in G_i(t)} l(w, x_d, y_d)}{| D_V |},
\end{equation}
where $l(w, x_d, y_d)$ is the error for datapoint $d$, and $| D_V |$ is the number of datapoints. Note that the function $l$ may include regularization terms that aim to prevent model overfitting~\cite{mcmahan2017communication}.

Fog computing allows (\ref{eq:glob}) to be solved in a distributed manner: instead of computing the solution at the server $s$, we can leverage computations at any device $i$ with available resources. Below, we follow the commonly used federated averaging framework~\cite{wang2019adaptive} in specifying these local computations and the subsequent global aggregation by the server in each iteration, illustrated by device $n$ in Figure~\ref{fig:ML}. To avoid excessive re-optimization at each device, the local updating algorithm does not depend on $G_i(t)$. We adjust the server averaging to account for the amount of data each device processes.

\subsubsection{Local loss minimization}
To distributedly solve (\ref{eq:glob}), we first decompose it into a weighted sum of local loss functions
\begin{equation}
\label{eq:loc}
L_i(w_i | G_i) = \frac{\sum_{t = 1}^T \sum_{d \in G_i(t)} l(w, x_d, y_d)}{ | G_i |},
\end{equation}
where $G_i \equiv \cup_{t \leq T} G_i(t)$ denotes the set of datapoints processed by device $i$ over all times. The global loss (\ref{eq:glob}) is then $L(w | D_V) = \sum_i L_i(w | G_i) \left|G_i\right|/\left|D_V \right|$ if $\cup_i G_i = D_V$, i.e., all datapoints $d \in D_V$ are eventually processed.

Loss functions such as (\ref{eq:loc}) are typically minimized using gradient descent techniques \cite{mcmahan2017communication}. Specifically, the devices update their local parameter estimates at $t$ according to
\begin{equation}
\label{eq:locUpdate}
w_i(t) = w_i(t-1) - \eta(t) \nabla L_i(w_i(t - 1) | G_i(t)),
\end{equation}
where $\eta(t) > 0$ is the step size or learning rate, and $\nabla L_i(w_i(t - 1) | G_i(t)) = \sum_{d \in G_i(t)} \nabla l(w_i(t - 1), x_d, y_d) / |G_i(t)|$ is the gradient with respect to $w$ of the average loss of points in the current dataset $G_i(t)$ at the parameter value $w_i(t - 1)$. We define the loss only on $G_i(t)$ since future data in $G_i$ has not yet been revealed; since we assume each node's data is i.i.d. over time, $L_i(w_i(t - 1) | G_i(t))$ approximates the local loss $L_i(w_i | G_i)$. 
The computational cost $c_i(t)$ of processing datapoint $d$ is then the cost of computing the gradient $\nabla l(w_i(t - 1), x_d, y_d)$. If the local data distributions $\mathcal{D}_i$ are all the same, then all datapoints are i.i.d. samples of this distribution, and this process is similar to stochastic gradient descent with batch size $|G_i(t)|$.

\subsubsection{Aggregation and synchronization}
The aggregation server $s$ will periodically receive the local estimates $w_i(t)$ from the devices, compute a global update based on these models, and synchronize the devices with the global update. Formally, the $k$th aggregation is computed as
\begin{equation}
\label{eq:agg}
w(k) = \frac{\sum_i H_i(k\tau) \cdot w_i(k\tau)}{\sum_i H_i(k\tau)},
\end{equation}
where $\tau$ is the fixed aggregation period and $H_i(k\tau) = \sum_{t = (k - 1)\tau + 1}^{k\tau} |G_i(t)|$ is the number of datapoints node $i$ processed since the last aggregation. Thus, the update is a weighted average factoring in the sample size $H_i(t)$ on which each $w_i(t)$ is based. Intuitively, since the objective in (1) minimizes the empirical loss~\cite{wang2019adaptive,mcmahan2017communication}, nodes that process more data should be weighted more, as in~(\ref{eq:glob})-(\ref{eq:agg}).

\begin{figure}
\vspace{-0.1in}
\centering
\includegraphics[width = 0.45\textwidth]{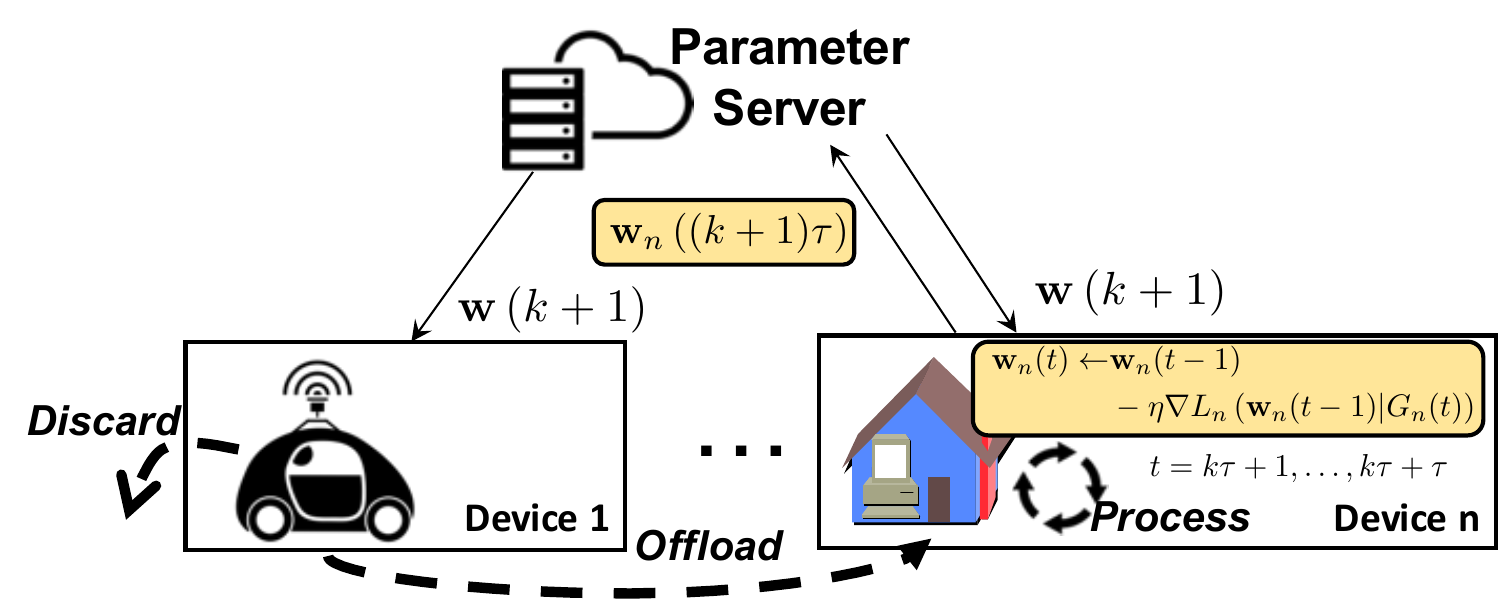}
\caption{Federated learning updates between aggregations $k$ and $k + 1$, in our system. Device 1 discards all of its data or offloads it to device $n$, which computes $\tau$ gradient updates on its local data. The final parameter values are averaged at the parameter server, with the result sent back to the devices to begin a new iteration.}
\label{fig:ML}
\vspace{-0.2in}
\end{figure}

Once this is computed, we synchronize local estimates, i.e., $w_i(t) \leftarrow w(t / \tau)$ $\forall i$. 
Using a smaller $\tau$ generally results in faster convergence of $w$, while larger $\tau$ requires less network resources. Prior work~\cite{wang2019adaptive} considered how to optimize $\tau$; we analyze its effect experimentally in Section \ref{sec:implementation}.

\vspace{-0.1in}
\subsection{Optimization Model for Data Processing Tasks}\label{subsec:formulation:opt}

We now consider the choice of $G_i(t)$, which defines the data processing task executed by device $i$ at time $t$. 
There are two possible reasons $G_i(t) \neq D_i(t)$: first, device $i$ may \emph{offload} some of its collected data to another device $j$ or vice versa, e.g., if $i$ has insufficient %does not have enough 
capacity to process all of the data $(D_i(t) \geq C_i(t))$\footnote{For notational convenience, $D_i(t)$ here refers to the number of datapoints $|D_i(t)|$, and similarly $G_i(t)$ refers to $|G_i(t)|$. The context will make the distinction clear throughout the paper.} 
or if $j$ has lower computing costs $(c_j(t) \leq c_i(t))$. Second, device $i$ may \emph{discard} data if processing it does not reduce the empirical loss (\ref{eq:glob}) by much. In Figure~\ref{fig:ML}, device 1 offloads or discards all of its data. 
We collectively refer to discarding and offloading as \emph{data movement}.

\subsubsection{Data movement model}
We define $s_{ij}(t) \in [0,1]$ as the fraction of data collected at device $i$ that is offloaded to device $j \neq i$ at time $t$. Thus, at time $t$, device $i$ offloads $D_i(t)s_{ij}(t)$ amount of data to $j$. Similarly, $s_{ii}(t)$ will denote the fraction of data collected at time $t$ that device $i$ also processes at time $t$. We suppose that as long as $D_i(t)s_{ij}(t) \leq C_{ij}(t)$, the capacity of the link between $i$ and $j \neq i$, then all offloaded data will reach $j$ within one time interval and can be processed at device $j$ in time interval $t + 1$. Since devices must have a link between them to offload data, $s_{ij}(t) = 0$ if $(i, j) \notin E(t)$.

We also define $r_i(t) \in [0,1]$ as the fraction of data collected by device $i$ at time $t$ that will be discarded. In doing so, we assume that device $j$ will not discard data that has been offloaded to it by others, since that has already incurred an offloading cost $D_i(t)s_{ij}(t)c_{ij}(t)$. The amount of data collected by device $i$ at time $t$ and discarded is then $D_i(t)r_i(t)$, and the amount of data processed by each device $i$ at time $t$ is
\begin{equation*}
G_i(t) = s_{ii}(t) D_i(t) + \sum_{j \neq i} s_{ji}(t-1)D_j(t-1).
\end{equation*}
In defining the variables $s_{ij}(t)$ and $r_i(t)$, we have implicitly specified the constraint $r_i(t) + \sum_j s_{ij}(t) = 1$: all data collected by device $i$ at time $t$ must either be processed by device $i$ at this time, offloaded to another device $j$, or discarded. We assume that devices will not store data for future processing, which would add another cost component to the model.

To quantify the cost of discarding relative to processing/offloading, we also define $f_i(t)$ as the cost per unit model loss $L(w_i(t) | D_V)$. The dependence on $i$ can capture the fact that certain devices in the network may weight their model error costs differently depending on the importance of the application. The dependence on $t$ can capture the fact that the loss may become less important relative to the resource utilization as the model approaches convergence.

\subsubsection{Data movement optimization}
We formulate the following cost minimization problem for determining the data movement variables $s_{ij}(t)$ and $r_i(t)$ over the time period $T$:
\begin{align}
& \underset{s_{ij}(t), r_i(t)}{\text{minimize}}
& & \sum_{t = 1}^T  \Bigg( \sum_i G_i(t) c_i(t) + \sum_{(i,j) \in E(t)} D_i(t) s_{ij}(t) c_{ij}(t)  \nonumber \\
& & & \quad \quad  + \sum_i f_{i}(t)L\left(w_i(t) | D_V \right)\Bigg), \label{eq:dataOpt} \\
& \text{subject to}
& & G_i(t) = s_{ii}(t) D_i(t) + \sum_{j \neq i} s_{ji}(t-1)D_j(t-1), \label{eq:process} \\
& & & s_{ij}(t) = 0, \;\; (i, j) \notin E(t), j \neq i, \\
& & & r_i(t) + \sum_j s_{ij}(t) = 1, \;\; s_{ij}(t), r_i(t) \geq 0, \label{eq:dist} \\
& & & G_i(t) \leq C_i(t), \;\; s_{ij}(t) D_i(t) \leq C_{ij}(t). \label{eq:cap}
\end{align}
Constraints (\ref{eq:process}--\ref{eq:dist}) were introduced above and ensure that the solution is feasible. The capacity constraints in (\ref{eq:cap}) ensure that the amounts of data transferred and processed are within link and node capacities, respectively. 

The three terms in the objective (\ref{eq:dataOpt}) correspond to the processing, offloading, and error costs, respectively. We do not include the cost of communicating parameter updates to/from the server in our model; unless a device processes no data, the number of updates stays constant. By including these three cost terms, our formulation accounts for the tradeoff between model performance and resource utilization: 

\para{(i) Processing, $G_i(t) c_i(t)$:} This is the computing cost associated with processing $G_i(t)$ of data at node $i$ at time $t$.

\para{(ii) Offloading, $D_i(t) s_{ij}(t) c_{ij}(t)$:} This is the communication cost incurred from node $i$ offloading data to $j$. 

\para{(iii) Error, $f_{i}(t)L\left(w_i(t) | D_V \right)$:} This cost quantifies the impact of data movement on the local model error at each device $i$. As the model approaches convergence for large $t$, improvement in the error term $L(w_i(t))$ with respect to the variables $G_i(t)$ becomes limited. Thus, we may have $f_i(t)$ decrease over time to prioritize network costs at later time periods. 

Since $w_i(t)$ is computed as in (\ref{eq:locUpdate}), the model error term is an implicit function of $G_i(t)$. We include the error from \emph{each} device $i$'s local model at each time $t$, instead of considering the error of the final model, since devices may need to make use of their local models as they are updated (e.g., if aggregations are infrequent due to resource constraints~\cite{wang2019adaptive}).

If the local datasets are i.i.d. (i.e., each is representative of the global distribution), then discarding more data clearly increases the global loss, since less data is used to train the ML model. Offloading may also skew the local model if it is updated over a small $G_i(t)$. We can, however, upper bound the loss function $L(w_i(t))$ regardless of the data movement:
\begin{theorem}[Upper bound on local loss]
\label{thm:error}
If $L_i(w)$ is convex, $\rho$-Lipschitz, and $\beta$-smooth, if $\eta \leq \frac{1}{\beta}$, and if $L(w(T)) - L(w^{\star}) \geq \epsilon$ for a lower bound $\epsilon$, then, after $K = \lfloor t / \tau \rfloor$ aggregations at time $t$ with a period $\tau$ and defining the constant $\delta_i \geq ||\nabla L_i(w) - \nabla L(w)||$, the local loss will satisfy
\begin{equation}
\label{eq:loss-bound}
L(w_i(t)) - L(w^{\star}) \leq \epsilon_0 + \rho g_i(t - K\tau),
\end{equation}
where $g_i(x) = \frac{\delta_i}{\beta} ((\eta \beta + 1)^x - 1)$, which implies $g_i(t - K\tau)$ is decreasing in $K$, and $\epsilon_0$ is given by
\begin{equation*}
\frac{1}{t\omega\eta(2 - \beta\eta)}+\sqrt{\frac{1}{t^2 \omega^2 \eta^2 (2 - \beta\eta)^2} + \frac{Kh(\tau) + g_i(t - K\tau)}{t\omega\eta(1 - \beta\eta/2)}}.
\end{equation*}
\end{theorem}

\begin{IEEEproof}
The full proof is contained in Appendix A of the supplementary material. We define $v_{k}(t)$, $t \in \{(k-1)\tau, ..., k\tau\}$ as the parameters under centralized gradient descent updates, $\theta_k(t) = L(v_k(t)) - L(w^{\star})$, and assume $\theta_k(k\tau) \geq \epsilon$ as in  \cite{wang2019adaptive}. After lower-bounding $\frac{1}{\theta_{K+1}(t)} - \frac{1}{\theta_{1}(0)}$ and $\frac{1}{L(w_i(t)) - L(w^{\star})} - \frac{1}{\theta_{K+1}(t)}$, we can upper-bound $L(w_i(t)) - L(w^{\star})$ as
\begin{equation*} 
\left(t \omega \eta \big( 1 - \frac{\beta \eta}{2} \big) - \frac{\rho}{\epsilon^2} \big( Kh(\tau) + g_i(t - K\tau) \big)\right)^{-1} = y(\epsilon).
\end{equation*}
Then, we let $\epsilon_0$ be the positive root of $y(\epsilon) = \epsilon$. The result follows since either $\underset{k \leq K}{\min} \; L(v_k(k\tau)) - L(w^{\star}) \leq \epsilon_0$ or $L(w_i(t)) - L(w^{\star}) \leq \epsilon_0$; both imply (\ref{eq:loss-bound}).
\end{IEEEproof}

In Section~\ref{sec:analysis}, we will consider how to use Theorem~\ref{thm:error}'s result to find tractable forms of the loss expression that allow us to solve the optimization (\ref{eq:dataOpt}--\ref{eq:cap}) efficiently. 
Moreover, without perfect information on the device costs, capacities, and error statistics over the time period $T$, we cannot solve (\ref{eq:dataOpt}--\ref{eq:cap}) exactly, so we will propose methods for estimating them.

\section{Optimization Model Analysis}
\label{sec:analysis}

We turn now to a theoretical analysis of the data movement optimization problem (\ref{eq:dataOpt})--(\ref{eq:cap}). We discuss the choice of error cost and capacity values (Section \ref{subsec:analysis:parameters}), and then characterize the optimal solution under different choices of the error cost and assumptions on the fog networking scenario for the ML use cases outlined in Section~\ref{sec:intro} (Section \ref{subsec:analysis:optimal}).

\vspace{-0.1in}
\subsection{Choosing Costs and Capacities}\label{subsec:analysis:parameters}
We may not be able to reliably estimate the costs $c_{ij}(t)$, $c_i(t)$, and $f_i(t)$ or capacities $C_i(t)$ and $C_{ij}(t)$ in real time. Misestimations are likely in highly dynamic scenarios that use mobile devices, since the costs $c_{ij}(t)$ of offloading data depend on network conditions at the current device locations. 
Mobile devices are also prone to occasional processing delays called ``straggler effects''~\cite{neglia2019role}, which can be modeled as variations in their capacities. The error cost, on the other hand, decreases over time as the model parameters move towards convergence. Here, we propose and analyze network characteristic selection methods. Although these methods also rely on some knowledge of the system, we show in Section V that a simple time-averaging of historical costs and capacities suffices to obtain reasonable data movement solutions. 

\subsubsection{Choosing capacities} Overestimating $C_i(t)$ leads to the deferral of some data processing until future time periods, which may cause a cascade of processing delays. 
Misestimations of the link capacities have similar effects. Here, to limit delays due to overestimation, we formalize guidelines for the capacities in (\ref{eq:cap})'s constraints. As commonly done~\cite{farhat2016stochastic}, we assume that processing times on stragglers follow an exponential distribution $\exp(\mu)$ for parameter $\mu$.

For device capacities, we obtain the following result:

\begin{theorem}[Processing time with compute stragglers]
\label{thm:compute-exp}
Suppose that the service time of processing a datapoint at node $i$ follows $\exp(\mu_i)$, and that $c_{ij}(t)$, $c_i(t)$, $C_i(t)$ are time invariant. We can ensure the average waiting time of a datapoint to be processed is lower than a given threshold $\sigma$ by setting the device capacity $C_i$ such that $\phi(C_i) = \sigma\mu_i / (1 + \sigma\mu_i)$, where $\phi(C_i)$ is the smallest solution to the equation $\phi = \exp\left(-\mu_i(1 - \phi)/C_i\right)$, an increasing function of $C_i$.
\end{theorem}
\begin{IEEEproof}
We note that the processing at node $i$ follows a D/M/1 queue with arrival rate $G_i(t) \leq C_i$, and the result follows from the average waiting time in such a queue. Details are in Appendix B of the supplementary material. 
\end{IEEEproof}

For instance, $\sigma = 1$ guarantees an average processing time of less than one time slot, as assumed in Section~\ref{sec:formulation}'s model. 
Thus, Theorem~\ref{thm:compute-exp} shows that we can still (probabilistically) bound the data processing time when stragglers are present. 
As long as $G_i(t) \leq C_i(t)$, where $G_i(t)$ is defined based on any values of $s_{ij}(t)$ and $r_i(t)$, Theorem~\ref{thm:compute-exp} holds for any data offloading and discarding policy, including the one provided by the solution to our optimization (5)-(9).

Network link congestion encountered in transferring data may also delay its processing, which can be handled by choosing the network capacity $C_{ij}(t)$ analogously to Theorem~\ref{thm:compute-exp}.

\subsubsection{Choosing error expressions}
To solve the optimization (\ref{eq:dataOpt})--(\ref{eq:cap}), we also need an expression of the error objective $f_i(t) L(w_i(t)|D_V)$ in terms of our decision variables. As shown in Theorem~\ref{thm:error}, we can bound the local loss at time $t$ in terms of a gradient divergence constant $\delta_i \geq \left\|\nabla L_i(w) - \nabla L(w)\right\|$. The following in turn provides an upper bound for $\delta_i$ in terms of $G_i(t)$:

\begin{lemma}[Error convergence]
\label{lem:iid}
Suppose the data distributions $\mathcal{D}_i$ have finite second and third moments. Then there exists a constant $\gamma_i >0$ independent of $G_i(t)$ such that
\begin{equation} \label{eq:error_conv}
\delta_i \equiv \left\|\nabla L_i\left(w | G_i(t)\right) - \nabla L(w)\right\| \leq \frac{\gamma_i}{\sqrt{G_i(t)}} + \frac{\gamma}{\sqrt{|D_V|}} + \Delta,
\end{equation}
where $\Delta = ||\nabla L_i(w | D_i)-\nabla L(w | D)||$, $\gamma = \sum_{i=1}^{N} {\gamma_i}$, and $|D_V|$ is the total number of datapoints generated.
\end{lemma}
\begin{IEEEproof}
We can express $||\nabla L_i(w | G_i(t))-\nabla L(w)||$ as the sum of $||\nabla L_i(w | G_i(t))-\nabla L_i(w | D_i(t))||$, $||\nabla L(w | D)-\nabla L(w | D_V)||$, and $\Delta$. We bound $||\nabla L(w | D) - \nabla L(w | D_V)||$ above by $\gamma / \sqrt{|D_V|}$ using the central limit theorem. Since $\nabla L(w|D_V)$ is the average of $\nabla L(w,x_d,y_d)$, $\forall (x_d,y_d) \in D_V$, we can view $\nabla L(w,x_d,y_d)$ as $|D_V|$ samples from a distribution whose expected value is $\nabla L(w|D)$. We repeat this argument for $||\nabla L_i(w|G_i) - \nabla L_i(w|D_i)||$.
\end{IEEEproof}

The bound in Lemma~\ref{lem:iid} is likely to be a loose bound for non-i.i.d data, since, with data movement, the resulting distribution of data at device $i$ will be a mixture of the original distributions at each device. 
The value of $\Delta$ quantifies the degree to which local datasets $\mathcal{D}_i$ are non-i.i.d. across devices. When the local datasets are i.i.d., $\mathcal{D}_i = \mathcal{D}_j$ $\forall i,j$ and $\Delta = 0$. Since we take the generated device datasets $D_i(t)$ as given in the optimization, $\Delta$ is independent of our decision variables $s_{ij}(t)$ and $r_i(t)$, and thus independent of $G_i(t)$. As a result, only the term $\gamma_i / \sqrt{G_i(t)}$ from~(\ref{eq:error_conv}) is dependent on our decision variables. 
By substituting Lemma~\ref{lem:iid} into $g_i(t)$ in Theorem~\ref{thm:error}, we see that $L(w_i(t)) - L(w^{\star}) \propto \sqrt{G_i(t)^{-1}}$. Since the other time-dependent terms in $g_i(t)$ can be absorbed into $f_i(t)$, one choice of the error cost $f_i(t)L(w_i(t)|D_V)$ in~(\ref{eq:dataOpt}) is $f_i(t) \sqrt{G_i(t)^{-1}}$, capturing diminishing marginal returns in $G_i(t)$.

Since $\sqrt{G_i(t)^{-1}}$ is a convex function of $G_i(t)$, with this choice of error cost, (\ref{eq:dataOpt}--\ref{eq:cap}) becomes a convex optimization problem and can be solved relatively easily in theory. 
When the number of variables is large, however -- e.g., if the number of devices $n > 100$ with $T > 100$ time periods, as could be the case in the applications discussed in Section \ref{subsec:intro:scenarios} -- standard interior point solvers will be prohibitively slow \cite{wong2016efficiency}. In such cases, we may wish to approximate the error term with a linear function and leverage faster linear optimization techniques, i.e., to take the error cost as $-f_i(t)G_i(t)$ so the error decreases when $G_i(t)$ increases. 
This choice eliminates the diminishing marginal returns in $G_i(t)$. 
For learning tasks that are of a critical nature (e.g., object detection in smart vehicles~\cite{cisco-5giiot}), this may actually be more desirable in order to prioritize the trained model accuracy. Ultimately, in practice, the desired relationship between the resource and error costs in the optimization depends on the particular application.

We can further express $-f_i(t) G_i(t) = -f_i(t) (1 - r_i(t)) D_i(t) + f_i(t) D_i(t) \sum_{j \neq i} s_{ij}(t) - f_i(t) \sum_{j \neq i} s_{ji}(t-1) D_j(t-1)$. Since these last two terms are linear functions of $s_{ij}(t)$, they can alternatively be viewed as part of the transmission cost $\sum_{(i,j) \in E(t)} D_i(t) s_{ij}(t) c_{ij}(t)$ in \eqref{eq:dataOpt}. Specifically, if we redefine $c_{ij}(t) \leftarrow c_{ij}(t) + f_i(t) - f_j(t+1)$ in the objective, we can treat the error cost as $-f_i(t)D_i(t)[1 - r_i(t)]$, which is equivalent to minimizing $f_i(t)D_i(t)r_i(t)$, i.e., a cost proportional to the amount of data discarded at node $i$. The analytical results we present in Sec.~\ref{subsec:analysis:optimal} will use this form.

We can quantify the difference imposed by the linear approximation by considering the Taylor expansion of $h(x) = x^{-1/2}$, where $x$ represents the quantity of processed data, i.e., $G_i(t)$, at a device. Our linear approximation is equivalent to the first order Taylor approximation of $h(x)$, i.e. $h(x) \approx h(x_0) - \frac{1}{2}x_0^{-3/2} (x-x_0)$, with $x_0 = 0$. Taylor's theorem shows that the error in this first order expansion is proportional to $c^{-5/2} x^2$ for some $0 < c < x$, which is increasing in $x$~\cite{bouman1996unified,nadh2017taylor}. Thus, the approximation will have a larger difference when devices are processing more data. Later in Sec. \ref{ssec:baseline}, our experiments will show that there can be a practical advantage to modeling the discard cost as $f_i(t)D_i(t)r_i(t)$ without modifying the $c_{ij}(t)$.

\vspace{-0.1in}
\subsection{Optimal Task Distributions}\label{subsec:analysis:optimal}
Given a set of costs and capacities for the optimization (\ref{eq:dataOpt}--\ref{eq:cap}), we now characterize the optimal data movement solutions. We consider two formulations that admit closed-form solutions and elucidate the relationships between our costs and decision variables: (i) convex discard costs over a hierarchical, static topology (Theorem~\ref{thm:nonlinear}), and (ii) linear discard costs over a general, possibly dynamic topology (Theorem~\ref{thm:unconstrained}).

\subsubsection{General and dynamic network topology}
First, we consider the case of a general network topologies. In this case, we can derive a closed-form solution for the data processing variables under the linear error term $f_{i}(t) r_i(t) D_i(t)$:

\begin{theorem}[Data movement with linear discard cost]\label{thm:unconstrained}
Suppose $C_i(t) \geq D_i(t) + \sum_{j\in \mathcal{N}_i(t - 1)} D_j(t - 1)$ for each device $i$, i.e., its compute capacity always exceeds the data it collects as well as any data offloaded to it by $\mathcal{N}_i(t-1) = \{j: (j,i) \in E(t-1)\}$. Then, if the error cost is modeled as $f_i(t)D_i(t)r_i(t)$, the optimal $s_{ij}^\ast(t)$ and $r_i^\ast(t)$ will each be $0$ or $1$, with the following conditions for them being $1$ at node $i$:
\begin{equation}
\label{eq:soln-linear}
\begin{cases}
s_{ik}^\ast(t) = 1 & \mbox{if } c_{ik}(t) + c_k(t + 1) \leq \min\left\{f_i(t),c_i(t)\right\} \\
s_{ii}^\ast(t) = 1 & \mbox{if } c_i(t) \leq \min\left\{f_i(t),c_{ik}(t) + c_k(t + 1)\right\} \\
r_i^\ast(t) = 1 & \mbox{if } f_i(t) \leq \min\left\{c_i(t),c_{ik}(t) + c_k(t + 1)\right\}
\end{cases}
\end{equation}
where $k = \underset{j: j \neq i, (i,j) \in E(t)}{\arg\min} \{c_{ij}(t) + c_j(t + 1)\}$.
\end{theorem}
\begin{IEEEproof}
Since $r_i(t) + \sum_j s_{ij}(t) = 1$ in (\ref{eq:dist}), each datapoint in $D_i(t)$ is either discarded, offloaded, or processed at $i$. It is optimal to choose the option with least marginal cost.
\end{IEEEproof}
This theorem implies that with a linear discard cost, in the absence of resource constraints, all data will either be processed, offloaded to the lowest cost neighbor, or discarded.
Additionally, Theorem~\ref{thm:unconstrained} quantifies how the link costs $c_{ij}(t)$ affect the amount of data offloaded, allowing us to write the fraction of data offloaded by device $i$ as $1-s_{ii}(t) - r_i(t)$, which is $1$ if $ c_{ik}(t) +c_{k}(t+1) \leq \min\{ f_i(t),c_i(t) \}$ and $0$ otherwise. We later use this point in the proofs for Theorems~\ref{thm:cost_uniform} and~\ref{thm:complexity}. 

\parab{Fog use cases.} We next move to characterize solutions for specific fog use cases outlined in Section \ref{sec:intro}. Table~\ref{tab:usecases} summarizes the topologies of these four applications. Networks in smart factories have fairly static topologies, since they are deployed in controlled indoor settings. They also exhibit a hierarchical structure, with less powerful devices connected to more powerful ones in a tree-like manner with devices at the same level unable to communicate, as shown in Figure~\ref{fig:topologies}. Connected vehicles have a similar hierarchical structure, with sensors and vehicles connected to more powerful edge servers, but their architectures are more dynamic as vehicles are moving. Similarly, AR applications feature (possibly heterogeneous) mobile AR headsets connected to powerful edge servers. Applications that involve privacy-sensitive data may have very different, non-hierarchical topologies as the links between devices are based on trust, i.e., comfort in sharing private information. Since social relationships generally change slowly compared to ML model training, these topologies are relatively static. 

While the connected vehicles and AR settings require the generic topology treatment from Theorem~\ref{thm:unconstrained}, we can derive additional results for static hierarchical and social topologies:

\subsubsection{Static and hierarchical topologies}
In hierarchical scenarios, more powerful edge servers will likely always have sufficient capacity to handle all offloaded data, and they will likely have lower computing costs when compared to other devices. If the cost of discarding is linear, then from Theorem~\ref{thm:unconstrained}, sensors offload their data to the edge servers, unless the cost of offloading exceeds the difference in computing costs. 
We show in Section~\ref{sec:implementation} that the network cost can exceed the savings from offloading to more powerful devices; in such cases, devices process or discard their data instead.

\begin{table}
\centering
\renewcommand{\arraystretch}{1.2}
\begin{tabular}{ccc}
\hline
{\bf Use case} & {\bf Topology} & {\bf Dynamics} \\ \hline \hline
Smart factories~\cite{cisco-5giiot} & Hierarchical & Fairly static \\ \hline
Connected vehicles~\cite{cisco-5giiot} & Hierarchical & Rapid changes \\ \hline
\multirow{2}{*}{Augmented reality~\cite{chatzopoulos2017mobile}} & Hierarchical, & Rapid changes \\ 
& heterogeneous & possible \\ \hline
Privacy-sensitive~\cite{ieee-5g-health,neglia2019role} & Social network & Fairly static \\ \hline
\end{tabular}
\caption{Dominant characteristics of the four fog use cases.}
\label{tab:usecases}
\vspace{-0.2in}
\end{table}

When the cost of discarding is nonlinear, the optimal solution is less intuitive: it may be optimal to discard data if the additional processing cost outweighs the expected error reduction. If we consider multiple heterogeneous devices with static processing costs and data generation rates, each connected to a more powerful edge server over the same wireless network (and thus assumed with the same connectivity costs), we can find a closed form solution to our original optimization:
\begin{theorem}[Data movement with nonlinear error costs]
\label{thm:nonlinear}
Suppose that $n$ devices with static processing costs $c_i(t) = c_i$ and data generation rates $D_i(t) = D_i$ can offload to an edge server, indexed as $n + 1$. Assume that there are no resource constraints, $c_i > c_{n + 1}$, the costs $c_{ij}(t) = c_t$ of transmitting to the server are identical and constant, and the discard cost is given by $f_{i}(t)L(w_i(t)) = \gamma / \sqrt{G_i}$ as in Lemma~\ref{lem:iid}. Then, letting $s_i$ denote the fraction of data offloaded for node $i$, for $D_i$ sufficiently large, the optimal amount of data discarded is
\begin{equation}
\label{eq:ropt}
r_i^\ast = 1 - \frac{1}{D_i} \left( \frac{\gamma}{2c_i} \right)^{\frac{2}{3}} - s_i, \; \forall i.
\end{equation}
Given the optimal $r_i^\ast$, an optimal solution for $s_i^\ast$ is given by
\begin{equation}
\label{eq:sopt}
s_i^{\ast} = \frac{1}{\sum_j D_j} \left( \frac{\gamma}{2(c_{n+1} + c_t)} \right)^{\frac{2}{3}}, \; \forall i.
\end{equation}
\end{theorem}

\begin{IEEEproof}
The full proof is contained in Appendix C. There we note that in the hierarchical scenario, the cost objective (\ref{eq:dataOpt}) can be rewritten as $\sum_i (1 - r_i - s_i) D_i c_i + \sum_i s_i D_i (c_{n+1} + c_t) + \sum_i \frac{\gamma}{\sqrt{(1 - r_i - s_i)D_i}} + \frac{\gamma}{\sqrt{\sum_i s_i D_i}}$. Taking the partial derivatives with respect to $r_i$ and $s_i$, and noting that a large $D_i$ forces $r_i, s_i \in [0, 1]$ for each node gives the result.
\end{IEEEproof}
Intuitively, as the costs $c_i$ or $c_{n+1}$ increase, so should the amount of data discarded, as in Theorem~\ref{thm:nonlinear}. Theorem~\ref{thm:nonlinear} also shows that data is neither fully discarded nor offloaded, in contrast with Theorem 3, implying that convex error bounds lead to a more balanced distribution of data across nodes in the network. We do not include resource constraints in Theorem~\ref{thm:nonlinear} in order to focus on the effects of the compute and transmission costs $c_i$, $c_{n + 1}$, and $c_t$ on the optimal solution. In practice, an edge server would have nearly unlimited capacity $C_{n+1}$ to process the datapoints generated by the devices, as it is much more powerful, and often has access to a more reliable power source, than typical mobile devices. Intuitively, including the capacities $C_i$ and $C_t$ on the devices and links would further increase the amount of data discarded.

\subsubsection{Socially-defined topologies}
When networks are larger and have more complex topologies, we extrapolate from Theorem~\ref{thm:unconstrained}'s characterization of device behaviors to understand data movement in the network as a whole. Specifically, we consider a social topology in which edges between devices are defined by willingness to share data (Figure~\ref{fig:social}). In such cases, we assume $c_{ij}(t) = 0$ as nodes either trust each other to share data or do not; nodes that do not trust each other simply will not have an edge between them. We can find the fraction of devices that offload data, which allows us to determine the cost savings from offloading, when the error cost is linear:

\begin{figure}[t]
\vspace{-0.1in}
\centering
\includegraphics[width = 0.22\textwidth]{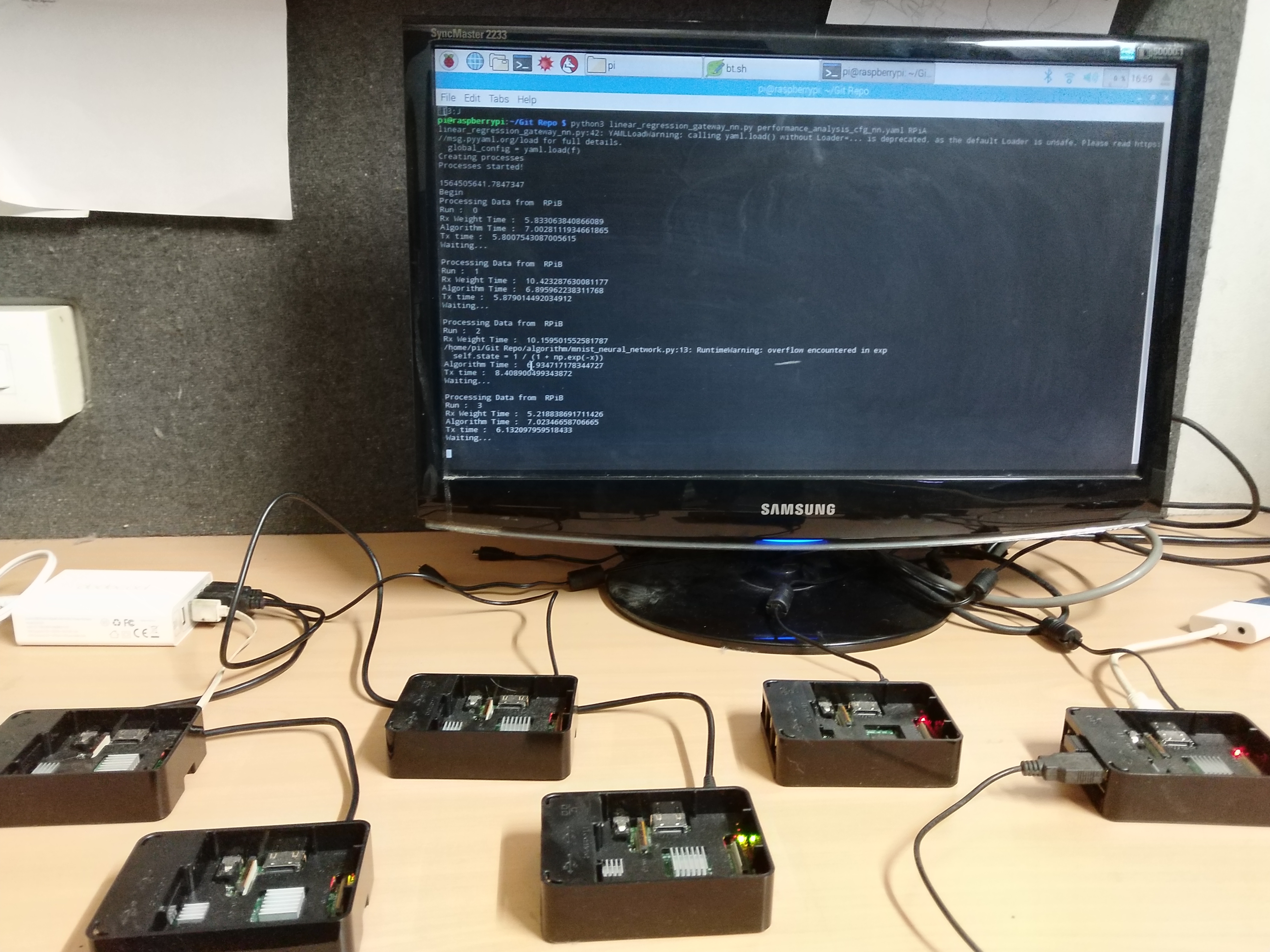}
\caption{Our Raspberry Pi devices running local computations.}
\label{fig:testbed}
\vspace{-0.2in}
\end{figure}

\begin{theorem}[Value of offloading]
\label{thm:cost_uniform}
Suppose the fraction of devices with $k$ neighbors equals $N(k)$. For a social network following a scale-free topology, for example, $N(k) = \Gamma k^{1-\gamma}$ for some constant $\Gamma$ and $\gamma\in(2,3)$. Suppose $c_i \sim U(0,C)$ and $c_{ij} = 0$ over all time, where $U(a,b)$ is the uniform distribution between $a$ and $b$ and no discarding occurs.

Then the average cost savings, compared to no offloading, equals
\begin{equation}
\label{eq:value}
\sum_{k = 1}^n N(k)\left(\frac{C}{2} - \frac{C(-1)^k}{k + 2} - \sum_{l = 0}^{k - 1} \binom{k}{l} \frac{C(-1)^l(k + 3)}{(l + 2)(l + 3)}\right).
\end{equation}
\end{theorem}
\begin{IEEEproof}
The full proof is contained in Appendix D of the supplementary materials. There, we use the result of Theorem 3 to find the probability that devices have lower processing cost neighbors, from which we can determine (\ref{eq:value}). 
\end{IEEEproof}

Thus, the reduction in cost from enabling device offloading in such scenarios is approximately linear in $C$: as the range of computing costs increases, there is greater benefit from offloading, since devices are more likely to find a neighbor with lower cost. The processing cost model may for instance represent device battery levels drawn uniformly at random from $0$ (full charge) to $C$ (low charge). The expected reduction from offloading, however, may be less than the average computing cost $C/2$, as offloading data to another device does not entirely eliminate the computing cost. Note we can use a similar technique to analyze the scenario where $c_{ij}$ is either nonzero or follows a given distribution; however, the expression for cost savings is then more complex due to the need to compare $c_j(t)$ (drawn from $U(0,C)$) with $c_{jk}(t) + c_k(t)$ (drawn from a convolution of $U(0,C)$ with the distribution of $c_{ij}(t)$). Similarly, we do not generalize Theorem~\ref{thm:cost_uniform}'s results to a non-uniform $c_i(t)$ distribution due to the complexity in writing a closed-form solution as in \eqref{eq:value}. However, we expect similar observations to hold, i.e., offloading becomes more beneficial as the cost range increases.

Next, we consider the case in which resource constraints are present, e.g., for less powerful edge devices. We find the expected devices that will have tight resource constraints:

\begin{theorem}[Probability of resource constraint violation]
\label{thm:complexity}
Let $N(k)$ be the number of devices with $k$ neighbors, and for each device $i$ with $k$ neighbors, let $p_k(n)$ be the probability that any one of its neighbors $j$ has $n$ neighbors. Also let $\tilde{C}$ denote the distribution of resource capacities, assumed to be i.i.d. across devices, and let $D_i(t) = D$ be constant. Then if devices offload as in Theorem~\ref{thm:unconstrained}, the expected number of devices whose capacity constraints are violated is
\begin{equation}
\hspace{-0.125in} \underset{\tilde{C}(x)}{\int} \left( \sum_{k = 1}^N N(k)\mathbb{P}\left[1 - P_o(k) + k\sum_{n = 1}^N\left(\frac{P_o(n)p_k(n)}{n}\right)\geq \frac{x}{D}\right] \right),
\label{eq:violation}
\end{equation}
with $P_o(k)$ defined as the probability a device with $k$ neighbors offloads its data based on the conditions in Theorem 3.
\end{theorem}
\begin{IEEEproof}
This follows from Theorem \ref{thm:unconstrained}, and from obtaining an expression for the expected amount of processed data at a node with $k$ neighbors when offloading is enabled. 
\end{IEEEproof}

Theorem~\ref{thm:complexity} makes two assumptions: $D_i(t) = D$ and resource capacities being i.i.d. across devices. The former assumption models a constant rate of data generation (e.g., from regular monitoring of user activity). If $D$ is stochastic (e.g., from event-driven sensor readings with random event sequences), we may generalize \eqref{eq:violation} by taking an additional expectation over $D$. The latter assumption is reasonable if we have no information about specific devices in the network, but know a range of possible hardware specifications that determine the compute capacities. This is likely to occur when there are a large number of devices; as we explain below, such a scenario is precisely where Theorem~\ref{thm:complexity} is most useful.

Theorem \ref{thm:complexity} allows us to quantify the complexity of solving the data movement optimization problem when resource constraints are in effect. We observe that it depends on not just the resource constraints, but also on the distribution of computing costs (through $P_o(k)$), since these costs influence the probability devices will want to offload in the first place. Additionally, Theorem 6 provides a guide for the most efficient way to solve optimization \eqref{eq:dataOpt}--\eqref{eq:cap}. If the expected number of resource violations is low, then following the procedure in Theorem 3 will produce a near-optimal solution that violates only a few resource constraints. We can then ensure these constraints are satisfied with minimal adjustments to the solution, e.g., optimizing over only the $s_{ij}(t)$ and $r_i(t)$ variables for the affected nodes and their neighbors while fixing all other optimization variables, or even increasing the $r_i(t)$ until the capacity constraints are satisfied. While this solution will not be optimal, \eqref{eq:soln-linear} can be implemented distributedly, if each device $j$ sends each of its neighbors $i$ (i) its processing cost $c_j(t)$ and (ii) estimates of $c_{ij}(t)$, e.g.,
channel conditions at the receiver. Thus, it is significantly more efficient than solving \eqref{eq:dataOpt}--\eqref{eq:cap} via a generic linear solver. On the other hand, if \eqref{eq:violation} is large, then we should solve \eqref{eq:dataOpt}--\eqref{eq:cap} using a generic optimizer.

\section{Experimental Evaluation}
\label{sec:eval}

\label{sec:implementation}

In this section, we experimentally evaluate our methodology. After discussing the setup in Sec.\ref{ssec:setup}, we investigate the performance of network-aware learning in Sec. \ref{ssec:baseline}. Then, we examine the effects of network characteristics, structure, and dynamics on our methodology in Sec. \ref{ssec:parameters} to \ref{ssec:movement}.

\begin{table}[t]
\begin{tabularx}{0.48\textwidth}{c *{4}{Y}}
\toprule[.2em]
\multirow{2}{*}{\bf{Learning Methodology}} & \multicolumn{2}{c}{\bf{Synthetic Costs}} & \multicolumn{2}{c}{\bf{Testbed Costs}} \\
\cmidrule(lr){2-3} \cmidrule(l){4-5}
& \bf{MLP} & \bf{CNN} & \bf{MLP} & \bf{CNN} \\
\midrule
Centralized & 92.00\% & 98.00\% & 92.00\% & 98.00\% \\
Federated (i.i.d.) & 90.82\% & 96.62\% & 90.82\% & 96.62\% \\
Federated (non-i.i.d.) & 84.00\% & 96.10\% & 84.00\% & 96.10\% \\
Network-aware (i.i.d.)& 88.63\% & 95.89\% & 89.70\% &  96.03\% \\
Network-aware (non-i.i.d.) & 83.20\% & 92.31\% & 85.47\% & 92.62\% \\
\bottomrule
\end{tabularx}
\caption{Comparison of accuracies obtained by learning methodologies for different cost settings and data distribution scenarios. Network-aware learning achieves within 4\% accuracy of federated learning on test datasets in all cases.} 
\label{tab: accuracy}
\vspace{-0.2in}
\end{table}

\vspace{-0.1in}
\subsection{Experimental Setup}
\label{ssec:setup}
\parab{Machine learning task and models.} We consider image recognition as our machine learning task, using the MNIST dataset~\cite{mnist}, which contains 70K images of hand-written digits labeled 0-9, i.e., a 10-class classification problem. We use 60K images as the training dataset $D_V$, and the remainder as our test set. The number of samples $|D_i(t)|$ at node $i$ is modeled using a Poisson arrival process with mean $|D_V|/(nT)$. For i.i.d. scenarios, each device $i$ generates $D_i(t)$ by sampling uniformly at random and without replacement from $D_V$. For non-i.i.d. scenarios, each device is limited to a random selection of five of the 10 possible labels. Device $i$ then samples $D_i(t)$ uniformly at random from this chosen subset of labels. Data distributions are i.i.d. unless stated otherwise. 
We train multilayer perceptrons (MLP) and convolutional neural networks (CNN) for image recognition on MNIST. We use cross entropy~\cite{lecun1989backpropagation} as the loss function $L(w | D_V)$, and a constant learning rate $\eta(t) = 0.01$. Unless otherwise stated, results are reported for CNN using $n = 10$ fog devices, an aggregation period $\tau = 10$, and $T = 100$ time intervals.

\parab{Network cost and capacity parameters.} To obtain realistic network costs for nodes $c_i(t)$ and links $c_{ij}(t)$, we collect measurements from a testbed consisting of six Raspberry Pis as nodes and AWS DynamoDB as a cloud-based parameter server (Figure~\ref{fig:testbed}). Three Pis collect data and transmit it over Bluetooth to another ``gateway'' Pi. The three gateway nodes receive this data and either perform a local gradient update or upload the data to DynamoDB. We measure 100 rounds of gradient update processing times and Pi-to-DynamoDB communication times while training a two-layer fully connected neural network (MLP), with devices communicating over 2.4 GHz WiFi or LTE cellular. These processing times are linearly scaled to range from 0 to 1, and recorded as $c_i(t)$, while the Pi-to-DynamoDB communication times are similarly scaled and saved as $c_{ij}(t)$. For completeness, we also evaluate performance for synthetic costs, where we take $c_{ij}(t), c_{i}(t) \sim U(0,1)$. Using both types of costs allows for a more thorough comparison between our network-aware and state-of-the-art learning frameworks, as network costs only affect network-aware learning. Unless otherwise stated, results are reported using the testbed-collected parameters.

\begin{figure}[t]
\vspace{-0.1in}
\hspace{-0.3in}
\centering
\captionsetup[subfigure]{oneside,margin={0.38cm,0cm}} 
\begin{subfigure}{0.24\textwidth}
\centering
\includegraphics[height=1.4in,trim = 3cm 8.5cm 1cm 8.2cm, clip]{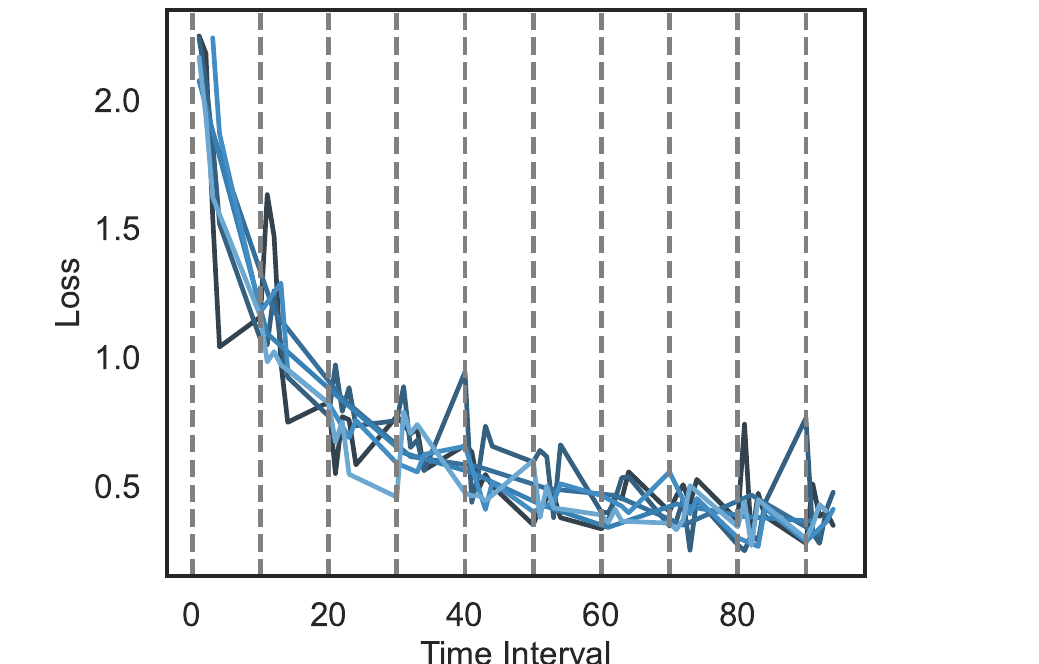}
\caption{Training loss over time}
\label{fig:loss}
\end{subfigure} 
\begin{subfigure}{0.24\textwidth}
\centering
\includegraphics[height=1.4in ,trim = 3cm 8.5cm 1cm 8.2cm, clip]{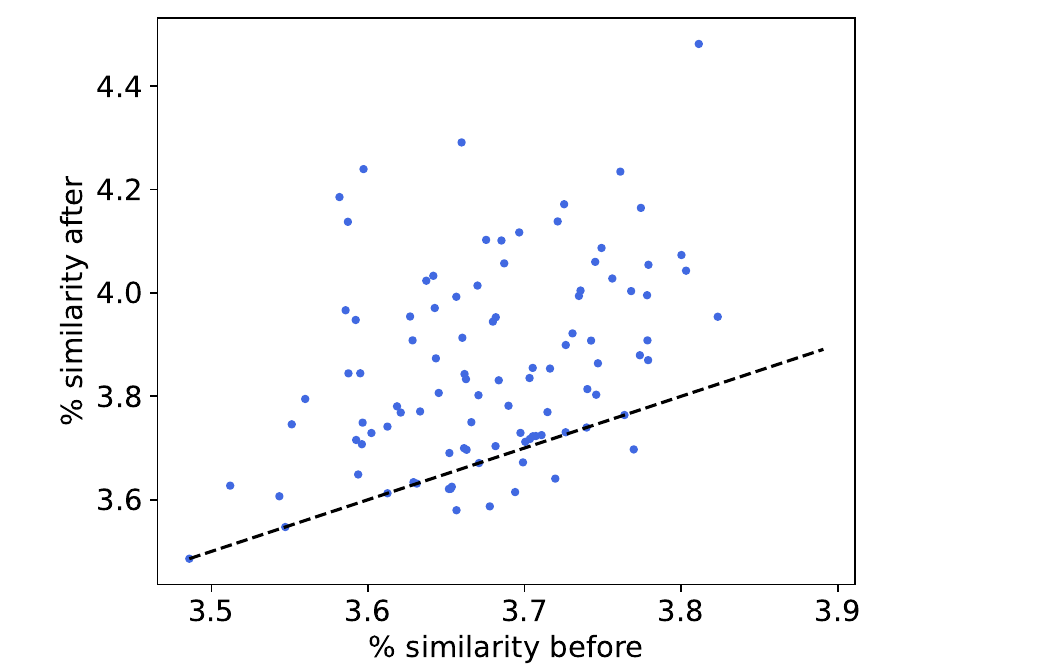}
\caption{Data similarity between devices}
\label{fig:data_similarity}
\end{subfigure}
\caption{(a) shows training loss over time for each device with network-aware learning. The average and variance drop over time. (b) captures average data similarity for 100 experiments across devices before ($x$-axis) and after ($y$-axis) offloading occurs for the non-i.i.d. scenario. The dotted line represents the change when no offloading occurs. 
Our methodology increases data similarity in almost all cases.}
\vspace{-0.2in}
\end{figure}

When imposed, the capacity constraints $C_i(t)$ and $C_{ij}(t)$ are taken as the average data generated per device in each time period, i.e., $|D_V|/(nT)$. The error costs $f_i(t)$ are modeled using the measurements from simulations on the Raspberry Pi testbed. All results are averaged over at least five iterations.

\parab{Centralized and federated learning.} 
To see whether our method compromises learning accuracy in considering network costs as additional objectives, we compare against a baseline of centralized ML training where all data is processed at a single device (server).
We also compare to federated learning with the same $\tau, T$ parameters where there is no data offloading or discarding, i.e., $G_i(t) = D_i(t)$.

\parab{Perfect information vs. estimation.} As discussed in Section \ref{subsec:analysis:parameters}, solving (\ref{eq:dataOpt}-\ref{eq:cap}) in practice requires estimating the costs and capacities over the time horizon $T$. To do this, we divide $T$ into $L$ intervals $T_1, ..., T_L$, and in each interval $l$, we use the time-averaged observations of $D_i(t)$, $c_i(t)$, $c_{ij}(t)$, and $C_i(t)$ over $T_{l-1}$ to compute the optimal data movement. The resulting $s^{\star}_{ij}(t)$ and $r^{\star}_i(t)$ for $t \in T_l$ are then used by device $i$ to transfer data in $T_l$. This ``imperfect information'' scheme will be compared with the ideal case in which the network costs and parameters are available (i.e., ``perfect information'').

\vspace{-0.1in}
\subsection{Efficacy of Network-Aware Learning}
\label{ssec:baseline}
We first investigate the efficicacy of our method on a fully-connected network topology  $E(t) = \{(i, j): i \neq j\}$. 

\subsubsection{Model accuracy} Table~\ref{tab: accuracy} compares the testing accuracies obtained by centralized, federated, and network-aware learning on both synthetic and testbed network costs. 
Our method achieves nearly the same (within 4\%) accuracy as federated learning in both the i.i.d. and non-i.i.d. cases. The performance differences between the i.i.d. and non-i.i.d. cases of each algorithm are expected, as non-i.i.d. local datasets are not representative of the overall dataset. 
Note that network-aware learning produced more accurate models on testbed rather than synthetic costs. In practical fog environments, outlined in Section~\ref{subsec:intro:scenarios}, devices with faster computations are also likely to transmit faster. The testbed-derived real costs contain such a correlation, which allows more cost-effective offloading and improves model accuracy.

The convergence of network-aware learning across individual devices is shown in Figure~4(a), with all devices exhibiting an overall decreasing trend. 
In Figure 4(b), we consider the similarity between local data distributions before and after offloading in the non-i.i.d. setting. Percent similarity is calculated between each pair of nodes $i$ and $j$ as the percent overlap in labels they contain -- i.e., $s_{i,j} = |\mathcal{Y}_i \cap \mathcal{Y}_j| / \min \{|\mathcal{Y}_i|, |\mathcal{Y}_j|\}$ where $\mathcal{Y}_i$ is the multiset of labels at device $i$ -- and then are averaged over all $i,j$ pairs. The average similarity increases due to offloading in nearly all cases, with an average improvement of around 10\%, which is a consequence of data offloading. 

\begin{table}[t]
\begin{tabularx}{0.484\textwidth}{c c c Y Y Y Y Y}
\toprule[.2em]
& \multicolumn{2}{c}{\bf{Accuracy (\%)}} & \multicolumn{5}{c}{\bf{Network costs (both i.i.d. and non-i.i.d.)}} \\
\cmidrule(lr){2-3} \cmidrule(lr){4-8}
& \bf{i.i.d.} & \bf{non-i.i.d.} & \bf{Process} & \bf{Transfer} & \bf{Discard} & \bf{Total} & \bf{Unit}\\
\midrule
A & 89.72 & 87.92 & 1234 & 0 & 0 & 1234 & 0.265 \\
B & 89.81 & 81.29 & 322 & 120 & 136 & 578 & 0.118 \\
C & 89.54 & 80.47 & 302 & 117 & 138 & 558 & 0.121 \\
D & 83.14 & 78.25 & 336 & 63 & 187 & 586 & 0.119 \\
E & 82.83 & 77.39 & 307 & 46 & 274 & 627 & 0.136 \\
\bottomrule
\end{tabularx}
\caption{Network costs and model accuracies obtained for i.i.d. and non-i.i.d. local data distributions in five different settings. The costs are the same for i.i.d. and non-i.i.d. local data distributions. The differences between A, where no data transfers are permitted, and B-E, which are variants of network-aware learning, show substantial improvements in resource utilization.}
\label{tab: cost}
\vspace{-0.2in}
\end{table}

\begin{figure*}[t]
\centering
\captionsetup[subfigure]{oneside,margin={0.75cm,0cm}}
\begin{subfigure}[t]{0.24\textwidth}
\centering
\includegraphics[height=1.4in]{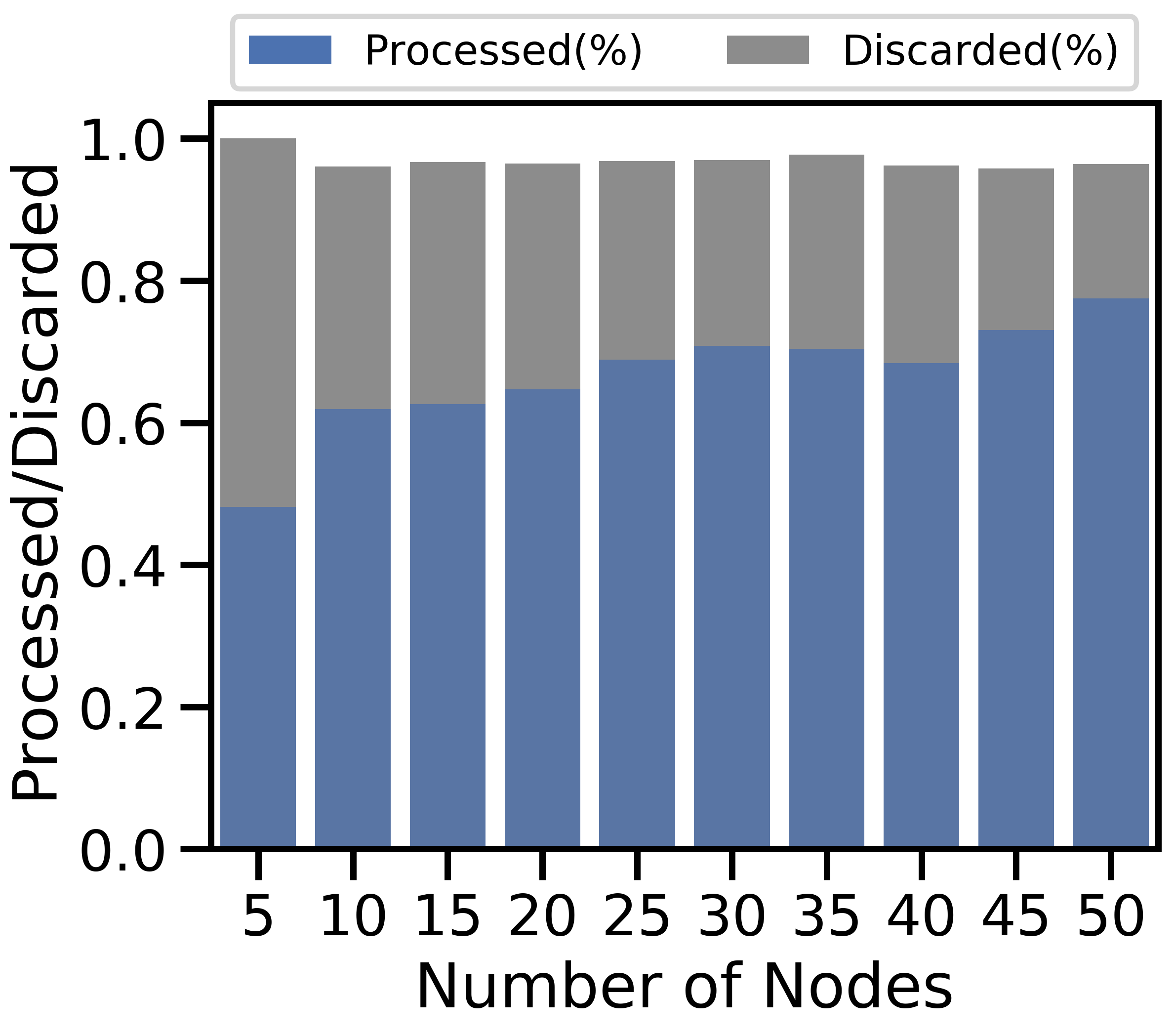}
\caption{Process vs. discard ratio}
\label{fig: varying_nodes_a}
\end{subfigure} 
\begin{subfigure}[t]{0.24\textwidth}
\centering
\includegraphics[height=1.4in]{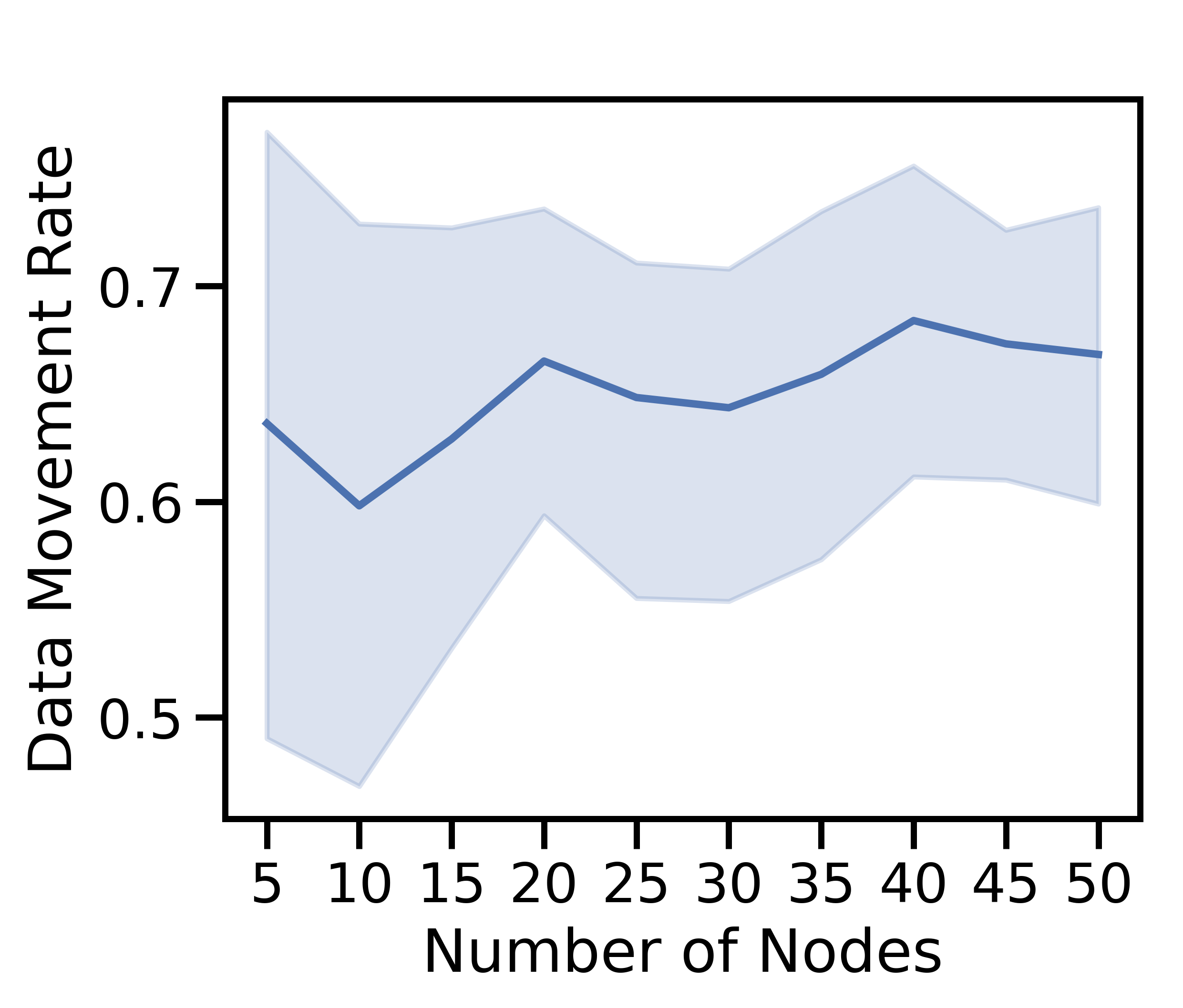}
\caption{Data movement rates}
\label{fig: varying_nodes_b}
\end{subfigure}
\begin{subfigure}[t]{0.24\textwidth}
\centering
\includegraphics[height=1.4in]{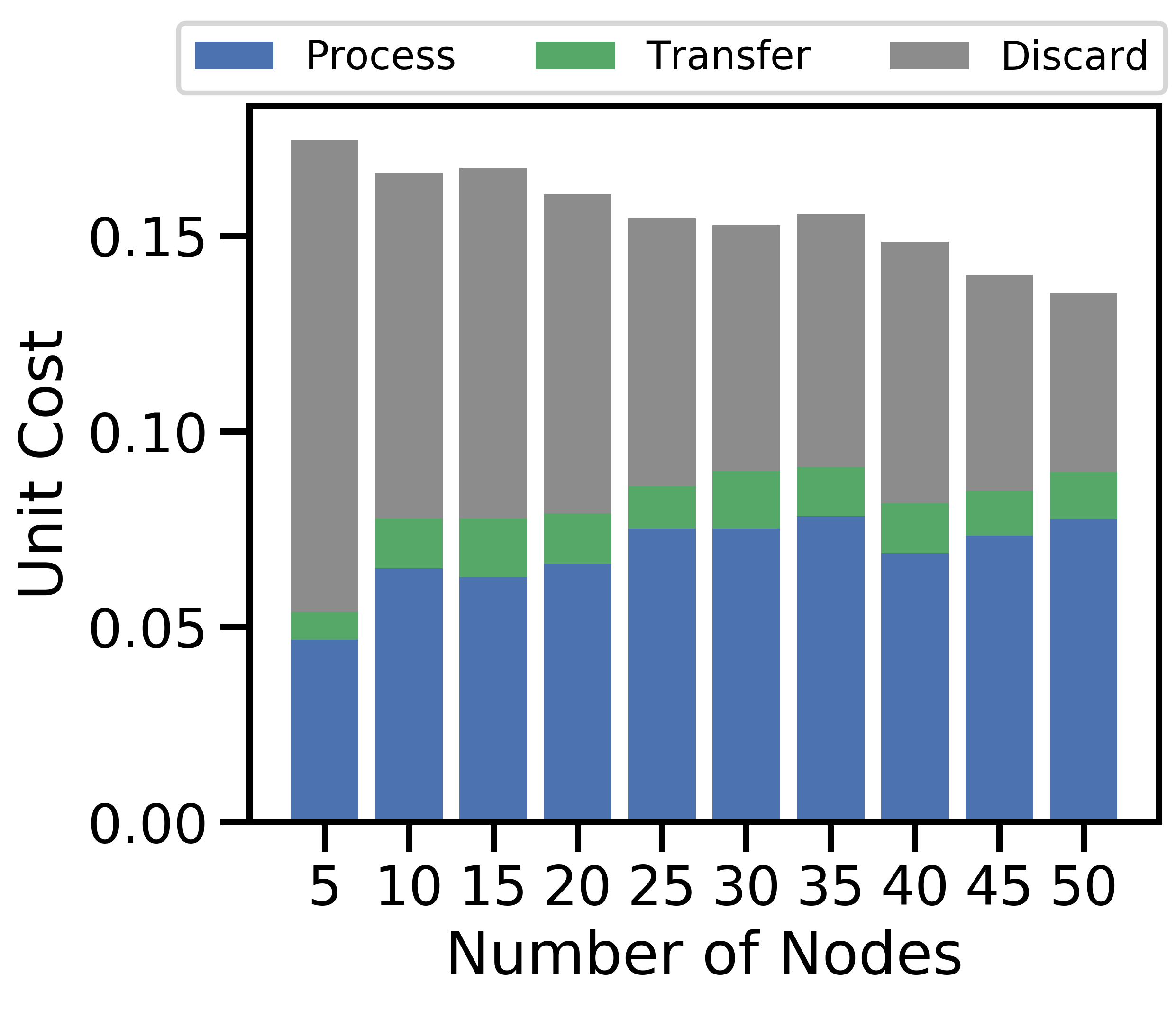}
\caption{Unit cost breakdown}
\label{fig: varying_nodes_c}
\end{subfigure}
\begin{subfigure}[t]{0.24\textwidth}
\centering
\includegraphics[height=1.4in]{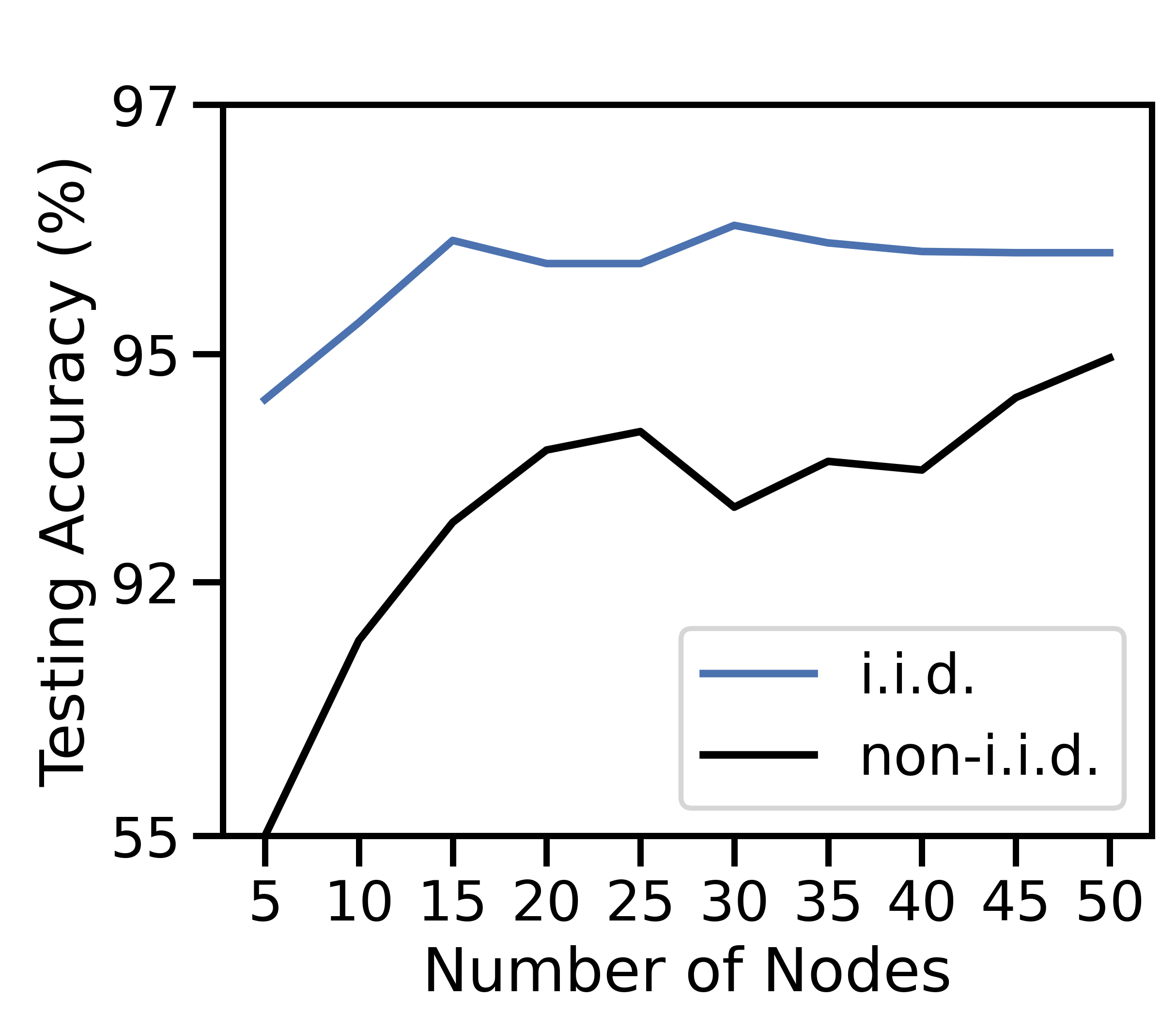}
\caption{Testing accuracy}
\label{fig: varying_nodes_d}
\end{subfigure}
\caption{Impact of the number of nodes $n$ on (a) the ratio of nodes processed vs. discarded, (b) the data movement rate, (c) the unit cost and cost components, and (d) the learning accuracy (log scale). The shading in (b) shows the range observed over time periods. We see that network-aware learning scales well with the number of nodes, as the cost incurred per datapoint improves, and the testing accuracy for the non-i.i.d. case improves substantially.}
\label{fig: varying_nodes}
\vspace{-0.15in}
\end{figure*}

\subsubsection{Offloading and imperfect information} While network-aware learning obtains similar accuracy to traditional federated learning, we expect it will improve network resource costs. Table \ref{tab: cost} compares the costs incurred and model accuracy for both the i.i.d. and non-i.i.d. scenarios in five settings, where settings B - E are applications of network-aware learning:
\begin{enumerate}[label = \Alph*.]
\item Offloading and discarding disabled
\item Perfect information and no capacity constraints
\item Imperfect information and no capacity constraints
\item Perfect information and capacity constraints
\item Imperfect information and capacity constraints
\end{enumerate}
Cost components in Table \ref{tab: cost} -- process, transfer, and discard (i.e., error) -- are summed over all nodes/links and time. The unit cost column is the total cost normalized by the total data generated in that setting, to account for variation in $D_i(t)$ across experiments. Simulations with imperfect information -- cases C and E -- allocate data based on historical network characteristics, which may be inaccurate, and can cause unintended large transfer or process costs. In the capacity limiting cases -- D and E -- the excess data must be discarded instead. Note also that the costs are the same for both the i.i.d. and non-i.i.d. settings as our offloading optimization (5)-(9) does not account for local data distributions.
Comparing A and B, we see that offloading reduces the unit cost by 53\%. The network takes advantage of transfer links, reducing the total processing cost by 74\%, by offloading more data. 
Our model is robust to estimation errors, similar to our observations from Section~\ref{subsec:analysis:parameters}, as we observe only minor changes in cost or accuracy from B to C, which has imperfect network information. 
When devices have strict capacity constraints, as in D and E, their gradient updates are based on fewer samples, and each node's $L_i(w_i(t))$ will tend to have larger errors.

\begin{table}[t]
\begin{tabularx}{0.488\textwidth}{c c c c *{4}{Y}}
\toprule[.2em]
& \bf{Discard cost} & \multicolumn{2}{c}{\bf{Accuracy~(\%)}} & \multicolumn{4}{c}{\bf{Network~costs}} \\ 
\cmidrule(lr){3-4} \cmidrule(lr){5-8}
& \bf{objective} & \bf{i.i.d.} & \bf{non-i.i.d.} & \bf{Pr} & \bf{Tr} & \bf{Di} & \bf{Tot}\\
\midrule
B & \multirow{2}{*}{$f_i(t) D_i(t) r_i(t)$} & 87.89 & 79.68 & 322 & 120 & 136 & 578 \\
D &  & 83.14 & 78.25 & 336 & 63 & 187 & 586 \\
\cmidrule(lr){1-8}
B & \multirow{2}{*}{$-f_i(t) G_i(t)$} & 90.05 & 80.90 & 390 & 461 & 125 & 976 \\
D &  & 87.86 & 78.08 & 410 & 244 & 136 & 790 \\
\cmidrule(lr){1-8}
B & \multirow{2}{*}{$f_i(t) / \sqrt{G_i(t)}$} & 86.81 & 78.04 & 323 & 79 & 172 & 574 \\
D &  & 85.42 & 77.26 & 311 & 83 & 184 & 578 \\
\bottomrule
\end{tabularx}
\caption{Effect of varying the discard cost model used in the optimization on the network costs and trained model accuracies, for settings B and D in Table~\ref{tab: cost}. Compared with the convex $f_i(t) / \sqrt{G_i(t)}$, the linear $-f_i(t) G_i(t)$ prioritizes higher accuracy over lower costs, while $f_i(t) D_i(t) r_i(t)$ is close to the convex case.}
\label{tab: comparison_convex_linear}
\vspace{-0.2in}
\end{table}

Overall, there is a roughly $7\%$ difference in accuracy for the i.i.d. case between settings A and E, and a 13\% difference for the non-i.i.d. case; this comes at an improvement of more than 50\% in network costs. The accuracy differences for the non-i.i.d. case are larger but exhibit the same trends across settings as the i.i.d. case. If higher accuracy is desired, a different error cost in~\eqref{eq:dataOpt} could be chosen, as we investigate next.

\begin{figure*}[t]
\centering
\captionsetup[subfigure]{oneside,margin={0.75cm,0cm}}
\begin{subfigure}[b]{0.24\textwidth}
\centering
\includegraphics[height=1.4in]{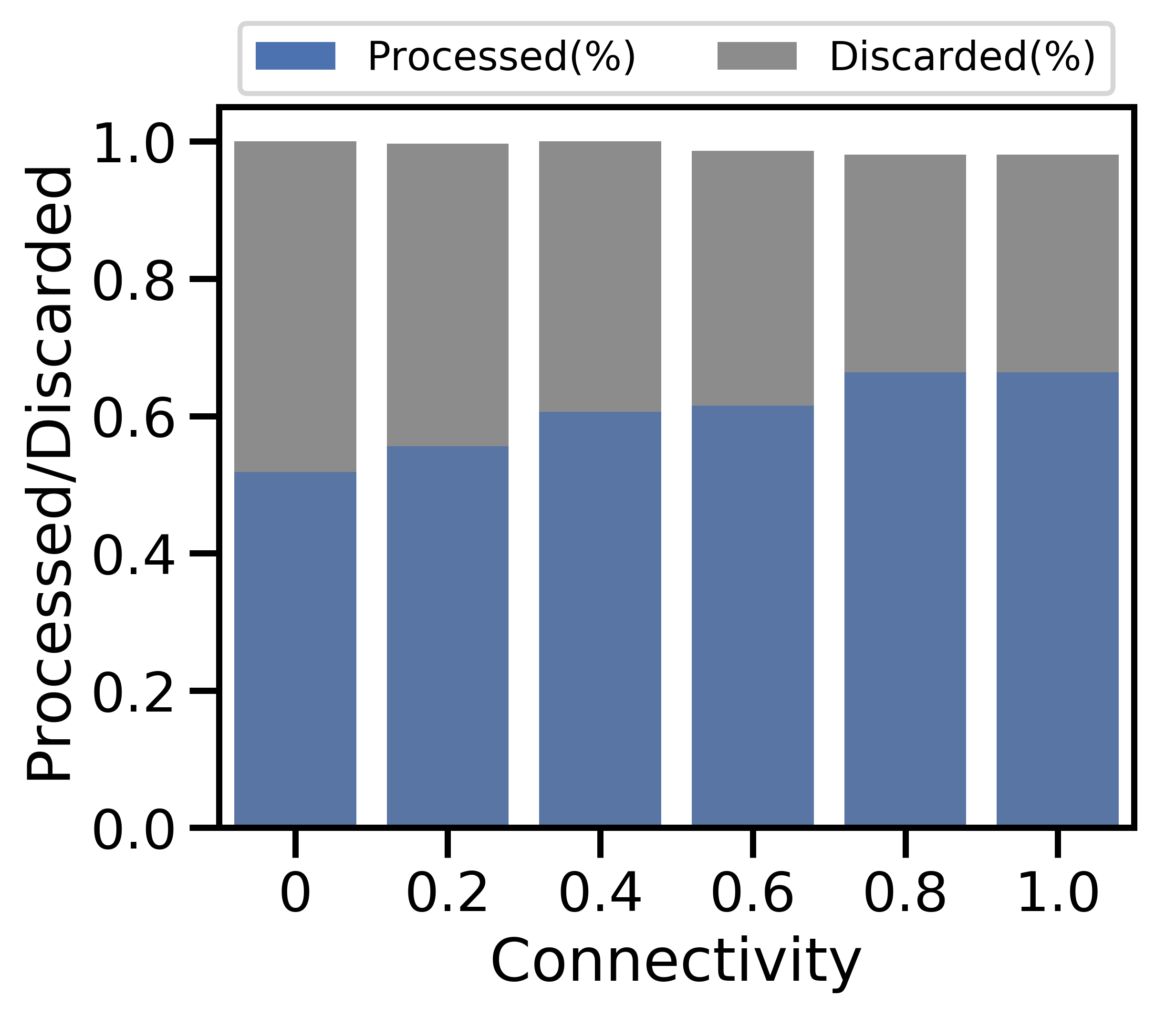}
\caption{Process vs. discard ratio}
\label{fig: varying_sparsity_a}
\end{subfigure} 
\begin{subfigure}[b]{0.24\textwidth}
\centering
\includegraphics[height=1.4in]{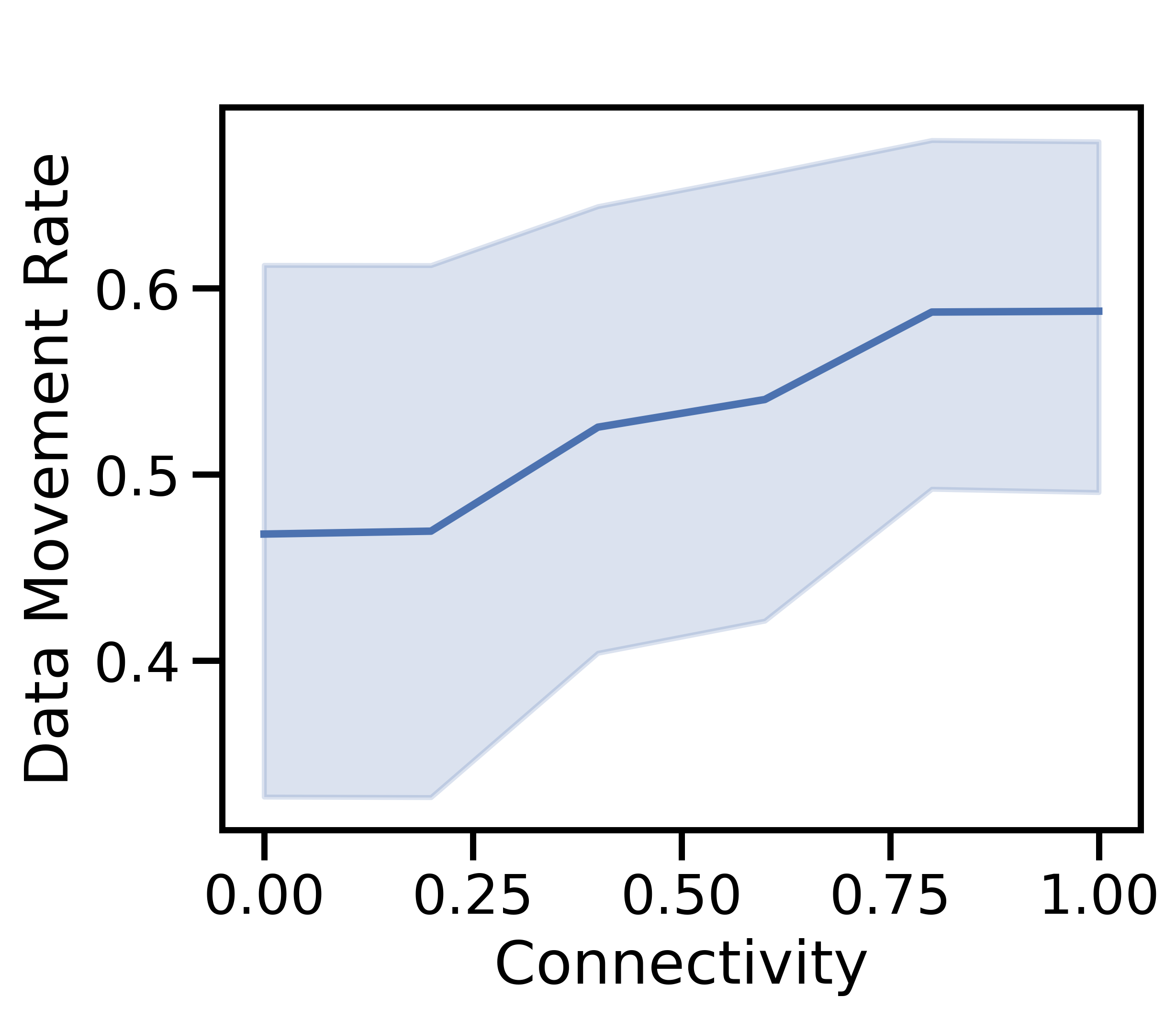}
\caption{Data movement rates}
\label{fig: varying_sparsity_b}
\end{subfigure}
\begin{subfigure}[b]{0.24\textwidth}
\centering
\includegraphics[height=1.4in]{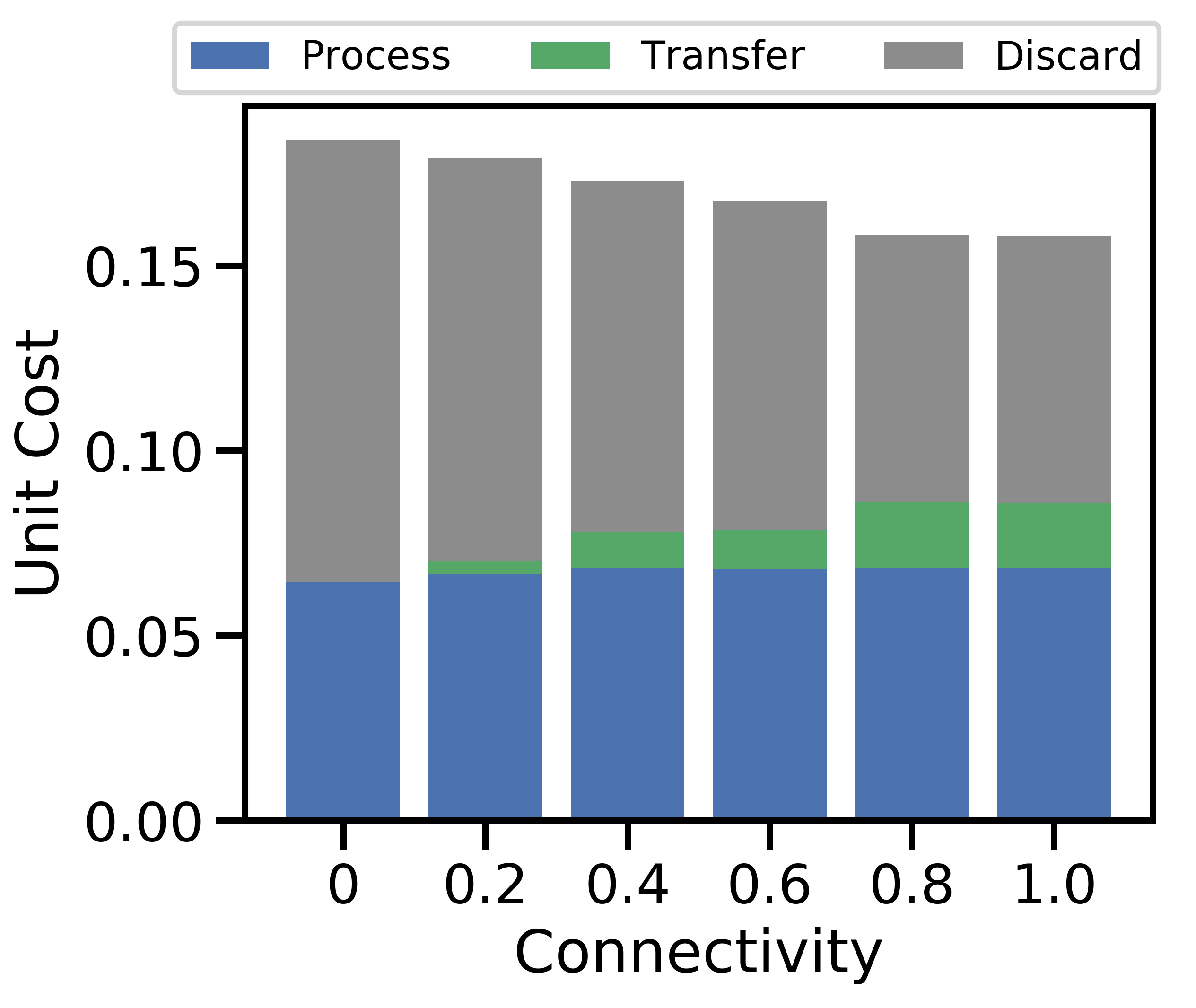}
\caption{Unit cost breakdown}
\label{fig: varying_sparsity_c}
\end{subfigure}
\begin{subfigure}[b]{0.24\textwidth}
\centering
\includegraphics[height=1.4in]{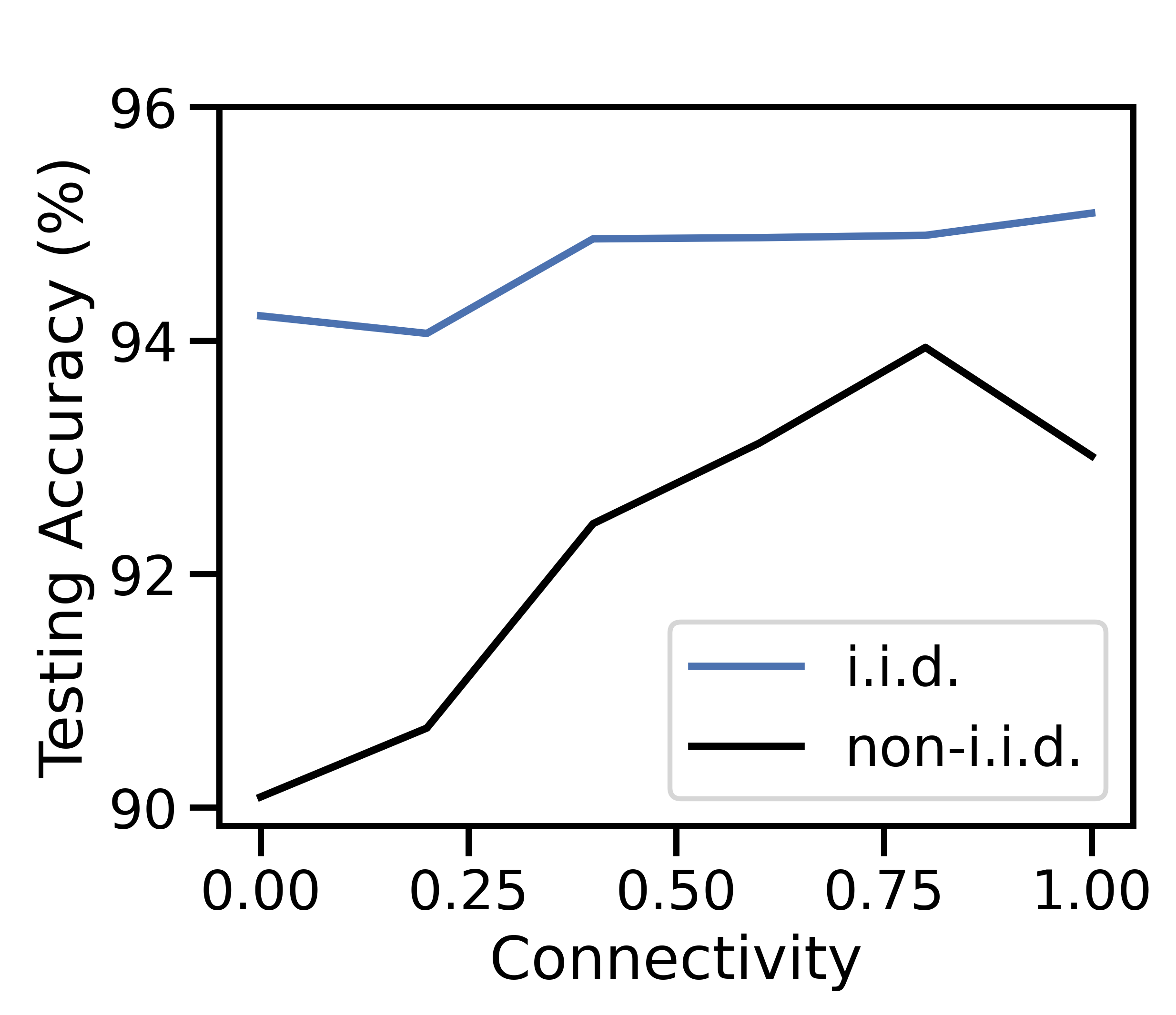}
\caption{Testing accuracy}
\label{fig: varying_sparsity_d}
\end{subfigure}
\caption{Impact of network connectivity $\rho$ on the different aspects of network-aware learning in Figure 5. Overall, we see that the costs have a linear relationship with $\rho$, that data movement rates increase in response to $\rho$, and that the learning accuracy tends to improve with $\rho$ particularly in the non-i.i.d. case. The shading in (b) indicates the range over time periods.}
\label{fig: varying_sparsity}
\vspace{-0.2in}
\end{figure*}

\subsubsection{Comparing error cost models}
\label{sssec:error}
We compare the use of three error cost models from Section~\ref{subsec:analysis:parameters}: $f_i(t) / \sqrt{G_i(t)}$, $-f_i(t) G_i(t)$, and $f_i(t) D_i(t) r_i(t)$ without modification to $c_{ij}(t)$. For this last case, note that $-f_i(t) D_i(t) (1 - r_i(t)) \geq -f_i(t) G_i(t)$ if $D_i(t) \sum_{j \neq i} s_{ij}(t) \leq \sum_{j \neq i} s_{ji}(t-1) D_j(t-1)$, as is likely since node $i$ is unlikely to offload more data at time $t$ than was offloaded to it at $t-1$. This upper bound on $-f_i(t) G_i(t)$ may work well for data-intensive applications as it neglects $f_i(t) / \sqrt{G_i(t)}$'s diminishing marginal returns. 

We show the results for the different cost models in Table~\ref{tab: comparison_convex_linear} under settings B and D from Table~\ref{tab: cost}. The linear cost $-f_i(t)G_i(t)$ produces a higher accuracy than using $f_i(t) / \sqrt{G_i(t)}$, but incurs a higher total cost due to transfer costs from offloading more data for processing. The results for $f_i(t) D_i(t) r_i(t)$ are close to the convex case. By neglecting the offloading terms in the discard cost, we prevent the solution from offloading more data to further reduce the error cost when it is exceeded by the marginal transfer cost.

\vspace{-0.1in}
\subsection{Effect of Network System Characteristics}
\label{ssec:parameters}
Our next experiments investigate the impact of the number of nodes, $n$, the network connectivity, $\rho$, and the aggregation period, $\tau$ on network-aware learning. 
We use a fully connected topology when varying $n$ and $\tau$, and, for varying $\rho$, we use a random graph with probability $P[(i,j) \in E(t), j \neq i] = \rho$. 

\subsubsection{Varying number of nodes $n$} Figure \ref{fig: varying_nodes} varies $n$ from $5$ to $50$ in increments of five nodes. Figure 5(a) depicts the fraction of data processed vs. discarded\footnote{Rounding used during the solution of the optimization problem resulted in the variance in the sum of the processed and discarded data ratios.}; 5(b) plots the change in the movement rate, i.e., fraction of data that is either offloaded or discarded; 5(c) breaks down unit cost by component; and 5(d) shows the model accuracy for i.i.d. and non-i.i.d.

Our method scales well as the unit cost in Figure 5(c) decreases with $n$. 
As the network grows, high-cost nodes are more likely to connect to low-cost nodes, which results in more offloading and is consistent with Theorems~\ref{thm:cost_uniform} and~\ref{thm:complexity}. 
Figure 5(b) confirms this: both the minimum and average data transfer rates grow with network size. As more offloading occurs, more data is processed in Figure 5(a). 
Although more data is processed, the increased processing cost is outweighed by the savings in discard cost in Figure 5(c). Training on more data then produces a more accurate ML model in Figure 5(d). 
In Figure 5(d), the non-i.i.d. case shows a substantial accuracy improvement as $n$ increases. Since each node in the non-i.i.d. case only contains half of the data labels for the ML problem, small networks are at a natural disadvantage compared to larger ones -- their empirical training dataset is unlikely to be representative of the underlying distribution, no matter what offloading scheme is used. 
Hence, we see the growth from 55\% to 95\% testing accuracy when $n$ increases from 5 to 50. 

\begin{figure*}[t]
\centering
\captionsetup[subfigure]{oneside,margin={0.75cm,0cm}}
\begin{subfigure}[t]{0.24\textwidth}
\centering
\includegraphics[height=1.4in]{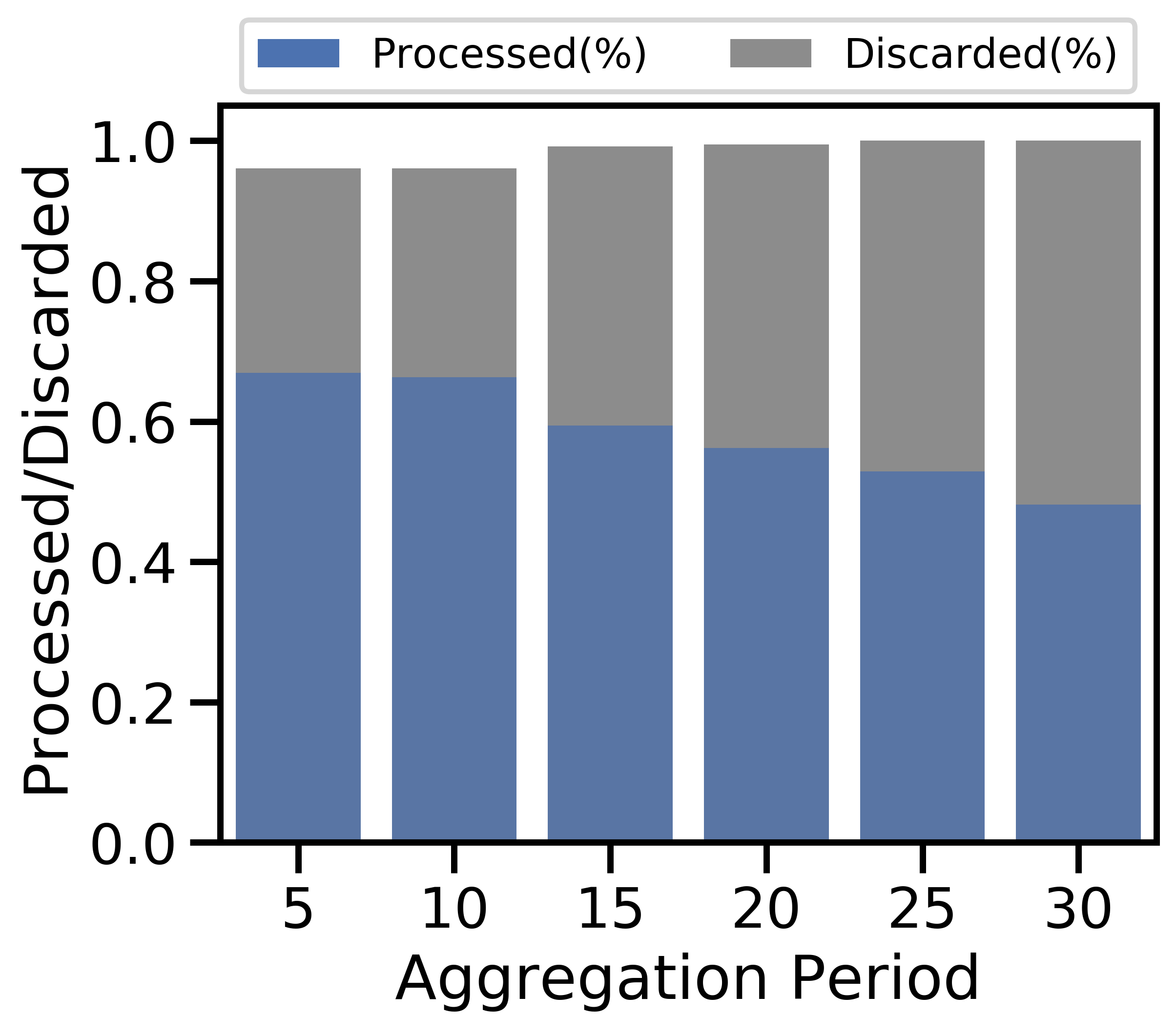}
\caption{Process vs. discard ratio}
\label{fig: varying_periods_a}
\end{subfigure} 
\begin{subfigure}[t]{0.24\textwidth}
\centering
\includegraphics[height=1.4in]{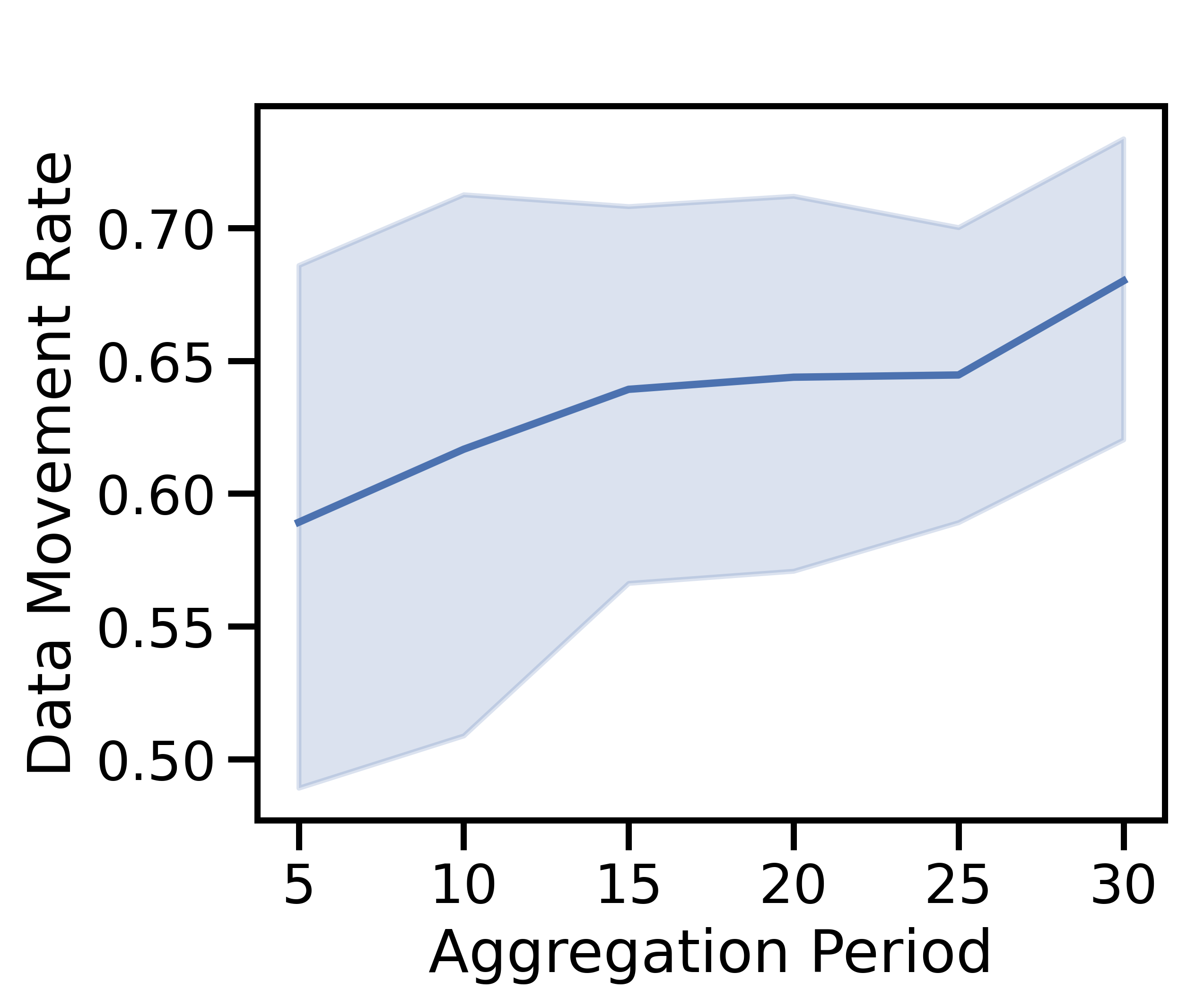}
\caption{Data movement rates}
\label{fig: varying_periods_b}
\end{subfigure}
\begin{subfigure}[t]{0.24\textwidth}
\centering
\includegraphics[height=1.4in]{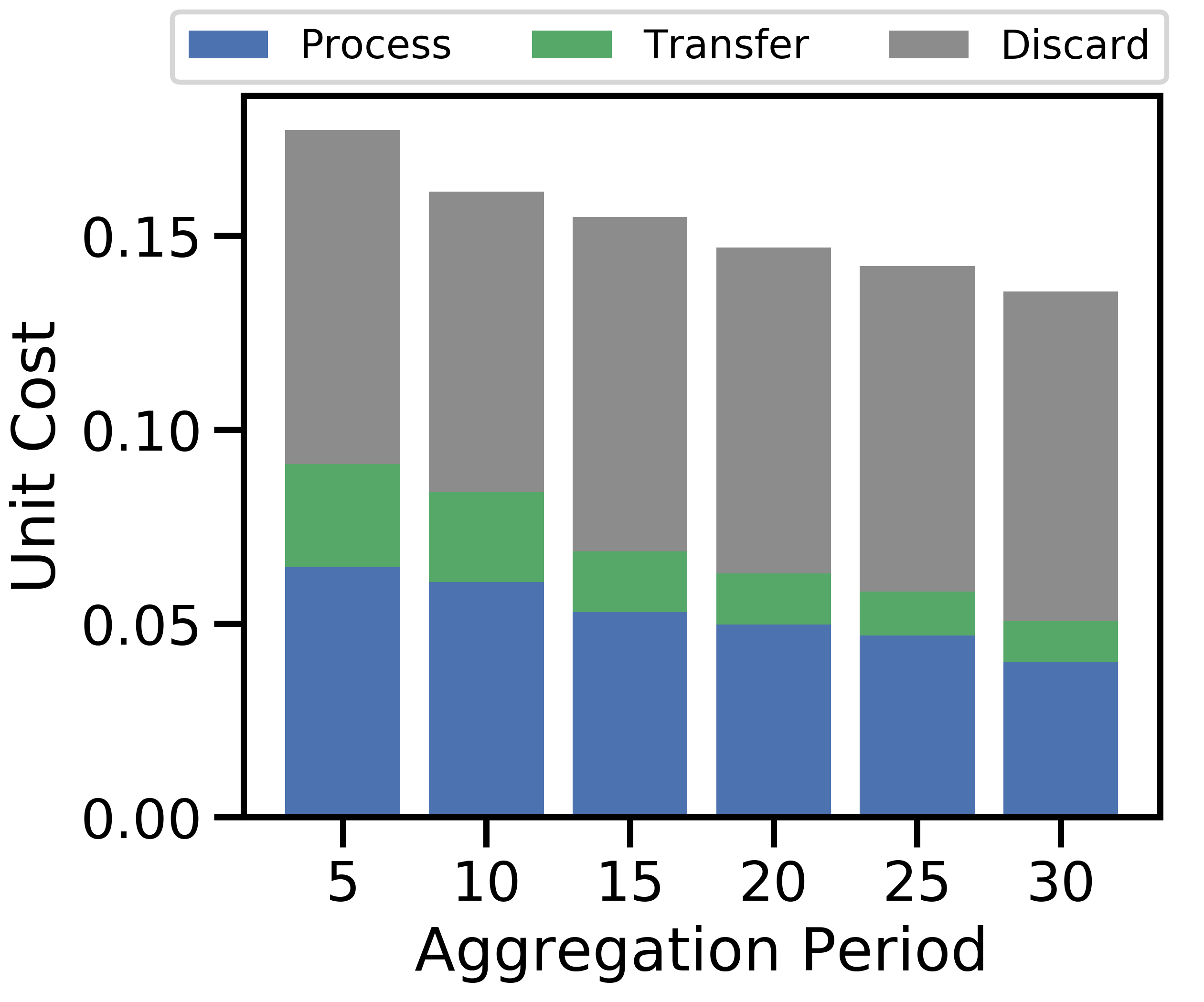}
\caption{Unit cost breakdown}
\label{fig: varying_periods_c}
\end{subfigure}
\begin{subfigure}[t]{0.24\textwidth}
\centering
\includegraphics[height=1.4in]{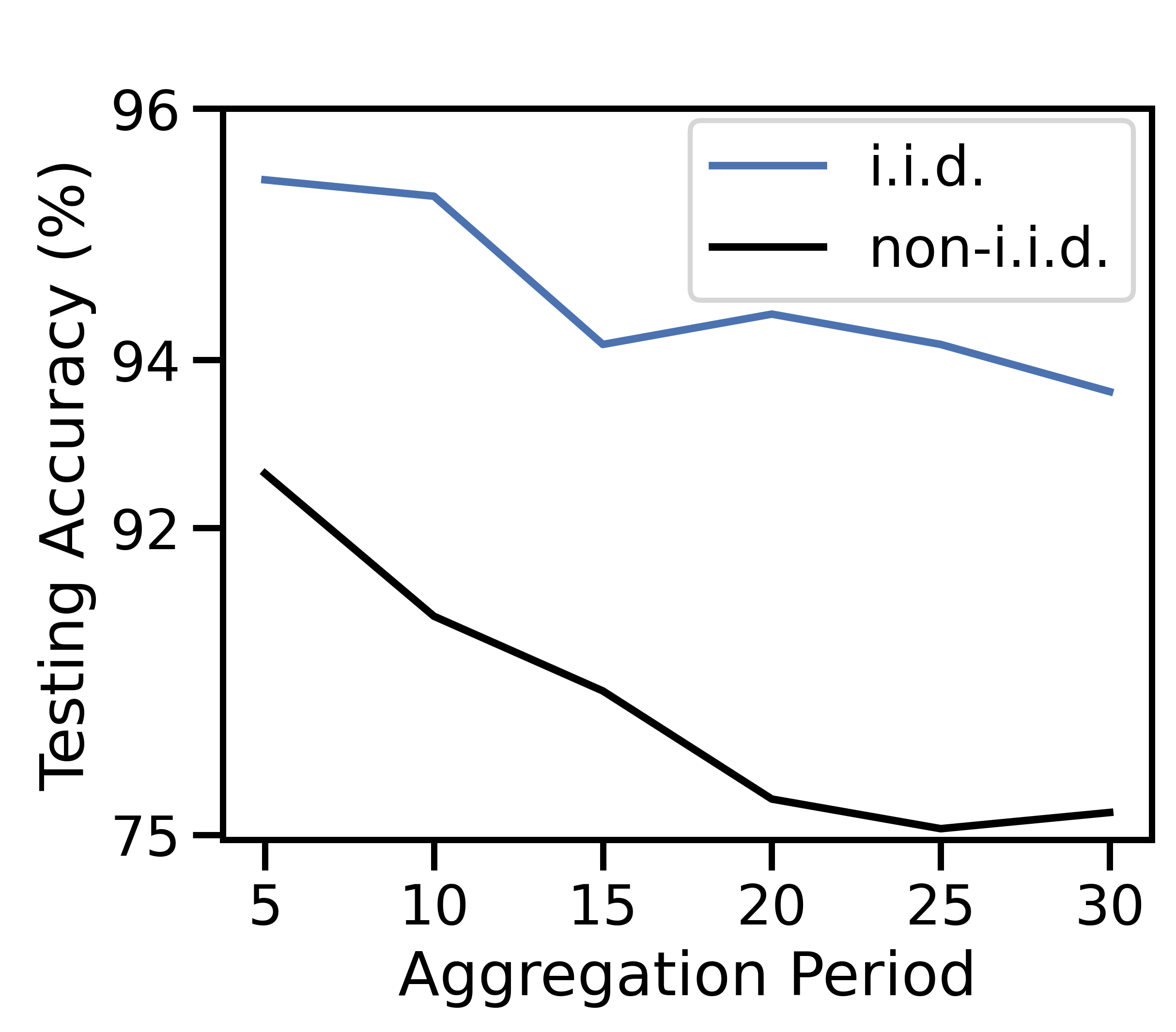}
\caption{Testing accuracy}
\label{fig: varying_periods_d}
\end{subfigure}
\caption{Impact of the aggregation period $\tau$ on the different aspects of network-aware learning in Figure 5. Overall, we see that a higher $\tau$ reduces total costs in (c), but decreases performance in (d), particularly for non-i.i.d., as the device models overfit their local distributions.}
\label{fig: varying_periods}
\vspace{-0.2in}
\end{figure*}

\subsubsection{Varying network connectivity $\rho$} Figure \ref{fig: varying_sparsity} examines the same characteristics in Figure \ref{fig: varying_nodes} as $\rho$ is varied from $0$ (i.e., completely disconnected) to $1$ (i.e., fully connected). Overall, we observe a similar trend to the effect of $n$: as connectivity grows, the unit cost per datapoint decreases in Figure 6(c), caused by cheaper alternatives to discarding. The relatively small change in total cost as $\rho$ varies indicates that network-aware learning is robust to variations in device connectivity.

High connectivity produces more opportunities for offloading in Figure \ref{fig: varying_sparsity}(b), which increases total data processed and decreases total data discarded in Figure \ref{fig: varying_sparsity}(a). Discard costs then take a smaller share of the unit costs in Figure \ref{fig: varying_sparsity}(c). Intuitively, more data processed leads to the more accurate model in Figure \ref{fig: varying_sparsity}(d). For non-i.i.d. data, the data transfer growth seen in Figures \ref{fig: varying_sparsity}(b) and \ref{fig: varying_sparsity}(c) increases the dataset similarity among devices, allowing model accuracy to improve. Increased network connectivity has a similar effect to having a larger network: there are more network links to nodes with low processing cost, which network-aware learning can leverage to produce cost savings without compromising model accuracy.

\subsubsection{Varying aggregation period $\tau$} Figure \ref{fig: varying_periods} varies the aggregation period $\tau$. Overall, a larger $\tau$ decreases unit cost and testing accuracy. 
We expect local models to converge as $\tau$ grows, similar to the effects studied in~\cite{wang2019adaptive}. 
As devices train on their local datasets for longer, they begin to converge, reducing the value of processing data. As a result, discarding becomes cost-effective in Figure \ref{fig: varying_periods}(a), and the discard costs dominate in Figure \ref{fig: varying_periods}(c), with the data movement rate in Figure \ref{fig: varying_periods}(b) increasing due to this extra discarding. Finally, since $T$ is constant, a higher $\tau$ reduces the number of global aggregations, which results in decreased test accuracy in Figure \ref{fig: varying_periods}(d), consistent with our findings in Theorem 1. 
Frequent aggregations are more important in the non-i.i.d. case where aggregations are needed to prevent overfitting to devices' distinct local distributions. Hence, large $\tau$ results in significantly worse accuracy for non-i.i.d. data.

\vspace{-0.1in}
\subsection{Effect of Fog Topology}
\label{ssec:topology}
Next, we evaluate network-aware learning on three fog computing topologies: hierarchical and social network topologies as in Section \ref{subsec:analysis:optimal}, and a fully-connected topology in which all nodes are neighbors. 
The social network is modeled as a Watts-Strogatz small world graph \cite{chiang2012networked} with each node connected to $n / 5$ of its neighbors, and the hierarchical network connects each of the $n / 3$ nodes with the lowest processing costs to two of the $2n / 3$ remaining nodes, randomly. 

\begin{figure}
\vspace{-0.1in}
\centering 
\includegraphics[width = 0.375\textwidth,trim = 3cm 8.2cm 3cm 8.1cm, clip]{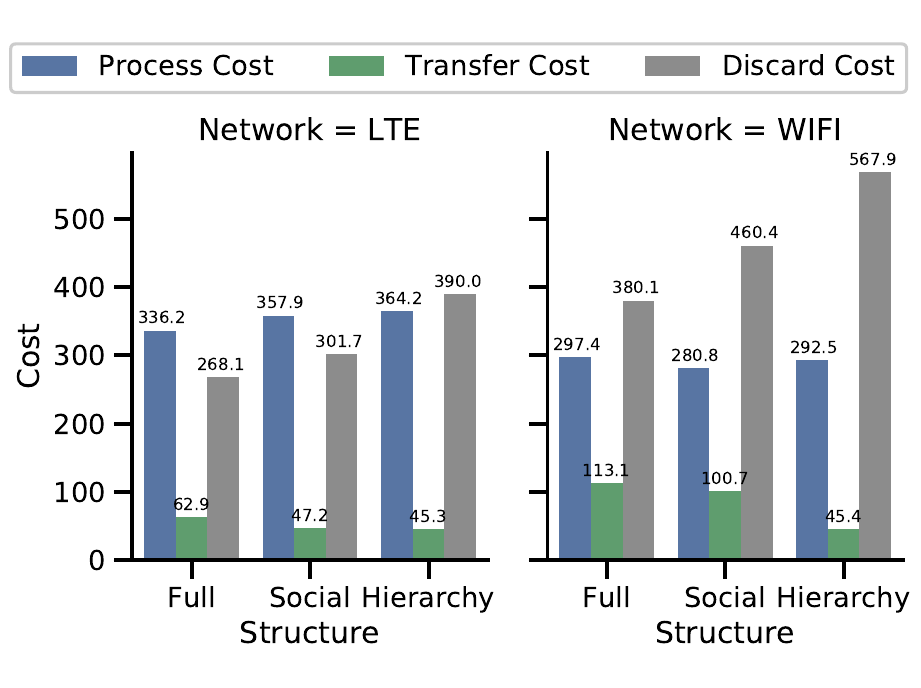}%eps}
\caption{Cost components for social, hierarchical, and fully connected topologies running network-aware learning on (a) LTE and (b) WiFi network media. Discard costs dominate for each topology in the case of WiFi, while the higher cost factor for LTE depends on the topology.}
\label{fig:topology}
\vspace{-0.2in}
\end{figure}

Our Raspberry Pi testbed provides LTE and WiFi network media for which we compare the network resource costs. 
Both media exhibit similar trends across the topologies in Figure \ref{fig:topology}. The topology determines offloading availability: the fully-connected topology maximizes the degree of each node, while the hierarchical topology minimizes the average degree. A smaller average degree limits offloading, which leads to more data processed locally and/or discarded. 
The major difference between LTE and WiFi is that WiFi skews more towards discarding. WiFi has fewer interference mitigation techniques than cellular, so, in the presence of several devices, we expect its links to exhibit longer delays. Consequently, both the discard and transfer costs are larger for WiFi than their LTE counterparts, regardless of topology. 
Results in both cases are consistent with the findings from varying the network connectivity in Figure \ref{fig: varying_sparsity} too: as networks become less connected and edges grow sparse, the ability of individual devices to offload their data to lower cost alternatives diminishes. Devices will marginally increase their data processing workloads, but ultimately a significant fraction of the data is discarded.

\vspace{-0.1in}
\subsection{Effect of Dynamic Networks}
\label{ssec:movement}
Finally, we consider network-aware learning when nodes may enter and exit the network. Initially, all devices are in the network. At each $t$, devices in the network will exit with probability $p_{exit}$, while devices outside the network will re-enter with probability $p_{entry}$. 
For worst-case analysis, nodes cannot transmit their local update results prior to exiting, and nodes rejoining the network cannot obtain the current global parameters until the ongoing aggregation period finishes. 

\begin{table}[t]
\begin{tabularx}{0.49\textwidth}{c c c*{5}{Y} }
\toprule[.2em]
\multirow{2}{*}{\bf{Setting}} & \multirow{2}{*}{\bf{Acc(\%)}} & \multirow{2}{*}{\bf{Nodes}} & \multicolumn{4}{c}{\bf{Cost}} \\
\cmidrule(lr){4-7}
& & & \bf{Process} & \bf{Transfer} & \bf{Discard} & \bf{Unit} \\
\midrule
Static & 95.83 & 10 & 399 & 66 & 328 & 0.135 \\ 
Dynamic & 94.79 & 7.8 & 300 & 56 & 256 & 0.144 \\ 
\bottomrule
\end{tabularx}
\caption{Comparison of network-aware learning characteristics on static and dynamic networks, with the probability of nodes entering and exiting fixed at 1\%. ``Acc'' represents model accuracy and ``Nodes'' represents the average number of active nodes per aggregation period. Overall, we see that node churn of 20\% has an impact of roughly 6\% on unit costs, and 1\% on accuracy.}
\label{tab: static_dynamic}
\vspace{-0.2in}
\end{table}

Table \ref{tab: static_dynamic} compares a dynamic network with $p_{exit} = p_{entry} = 1\%$ against the static case. 
In a dynamic network, network-aware learning operates with less overall network data and compute capability as active nodes per aggregation period decreases from 10 in the static case to an average of 7.8 in the dynamic case. 
Node exits always result in at least one inactive node - even if a new node enters, it must wait for the synchronized global parameters. 
So, our methodology is reasonably robust in dynamic networks as a 20\% decline in active nodes/period only leads to a 6\% increase in unit costs incurred per datapoint, and a 1\% accuracy decline, due to fewer processed data and more discarding (also visible from the ratio of processed to discard costs). 

\begin{figure*}[t]
\centering
\captionsetup[subfigure]{oneside,margin={0.1cm,0.1cm}} 
\begin{subfigure}[t]{0.178\textwidth} 
\centering
\includegraphics[height=1.3in,scale=0.9]{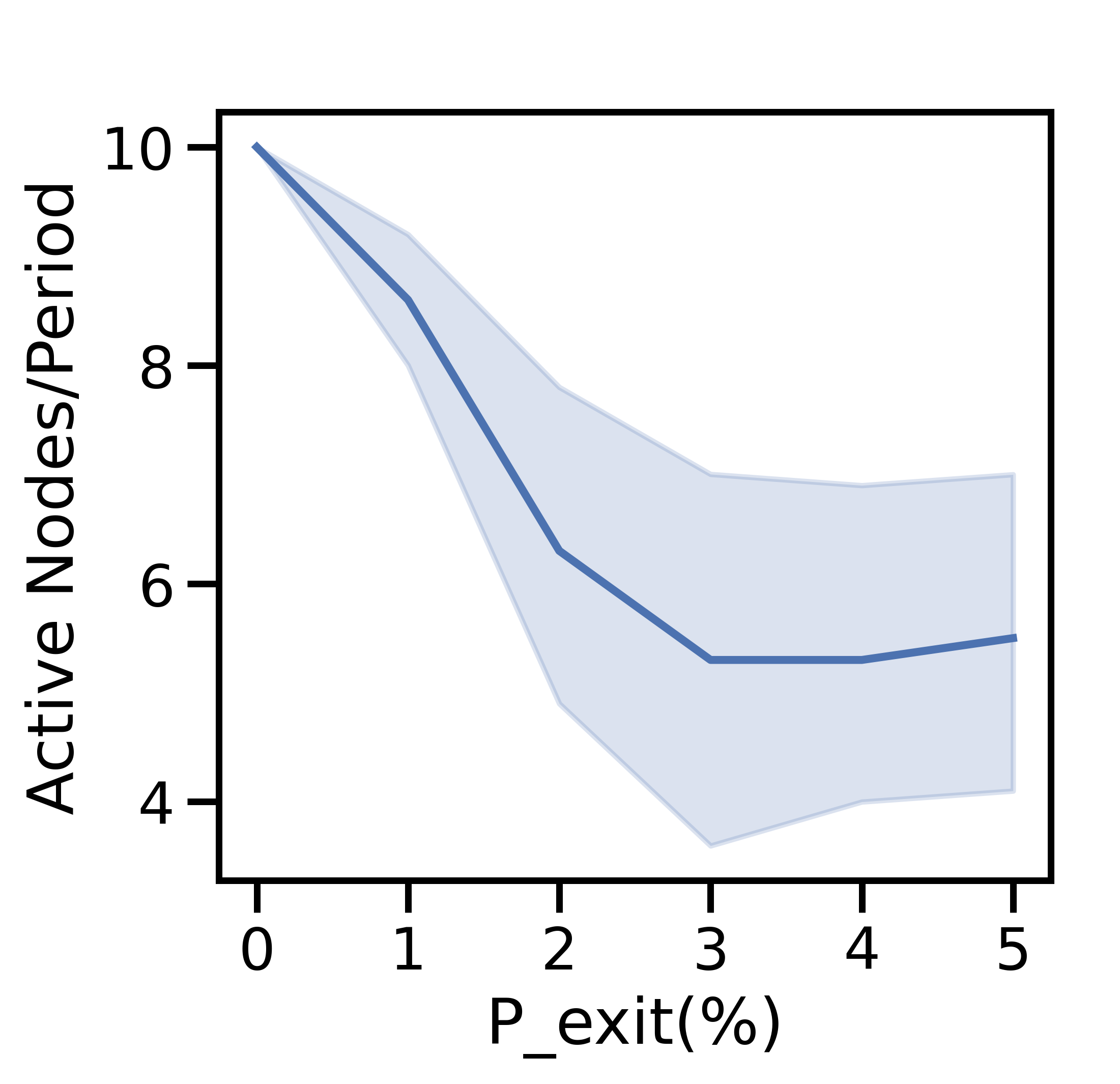}
\caption{Mean active nodes}
\label{fig: varying_dynamic_exit_e}
\end{subfigure}
\begin{subfigure}[t]{0.192\textwidth} 
\centering
\includegraphics[height=1.3in,scale=0.9]{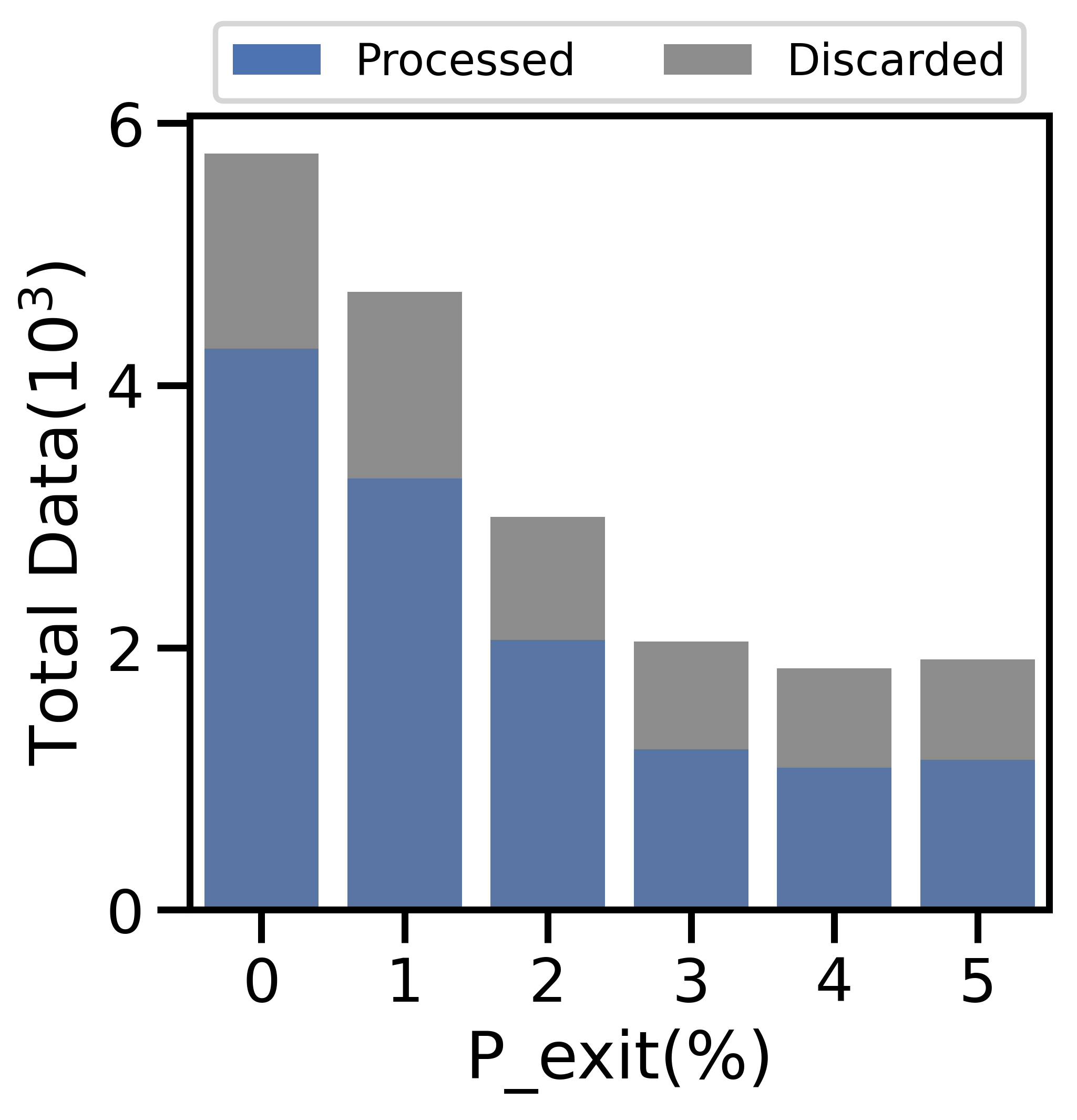}
\caption{Process vs. discard}
\label{fig: varying_dynamic_exit_a}
\end{subfigure} 
\begin{subfigure}[t]{0.20\textwidth} 
\centering
\includegraphics[height=1.3in,scale=0.9]{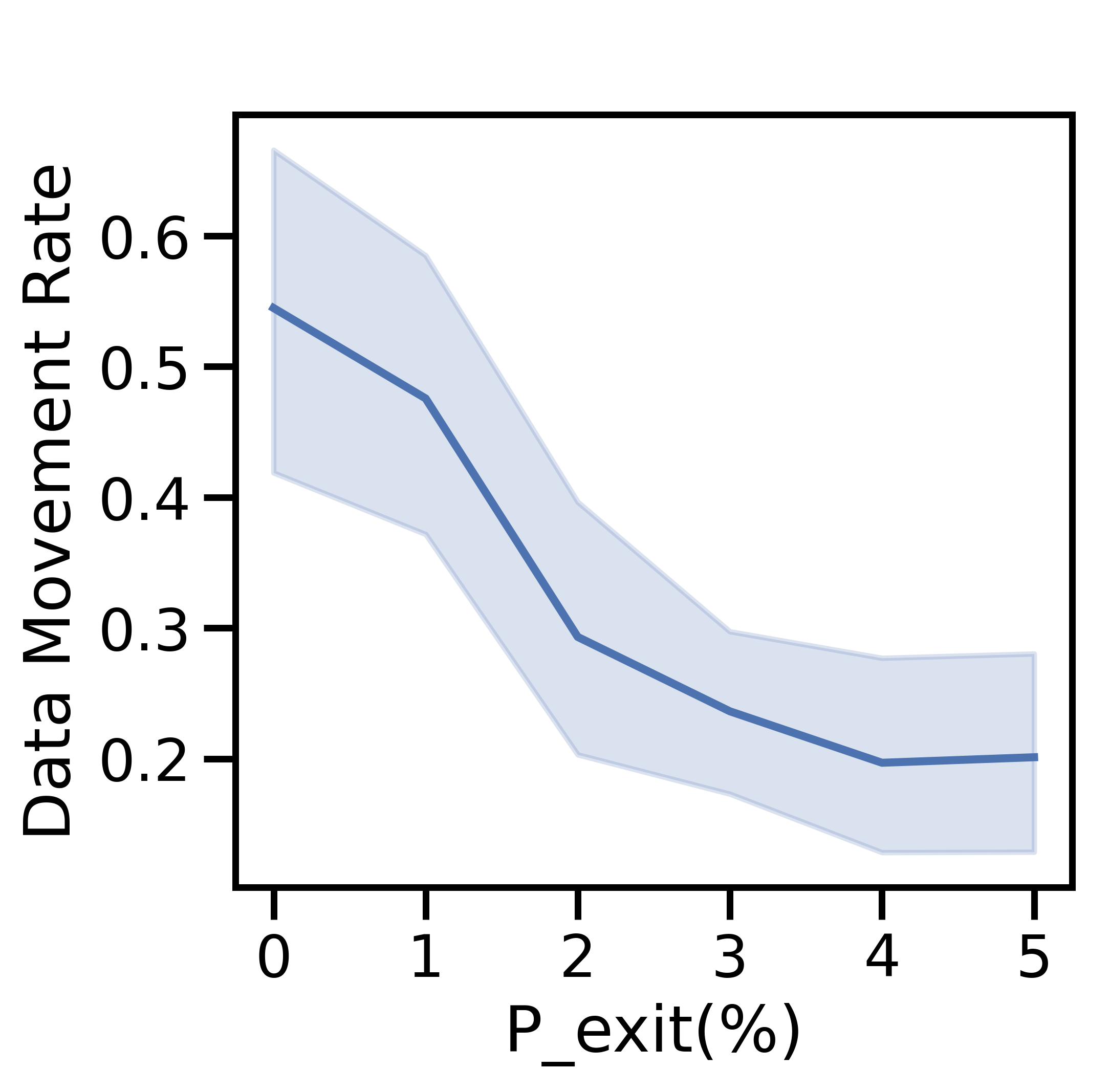}
\caption{Data movement rates}
\label{fig: varying_dynamic_exit_b}
\end{subfigure}
\begin{subfigure}[t]{0.2\textwidth} 
\centering
\includegraphics[height=1.3in,scale=0.9]{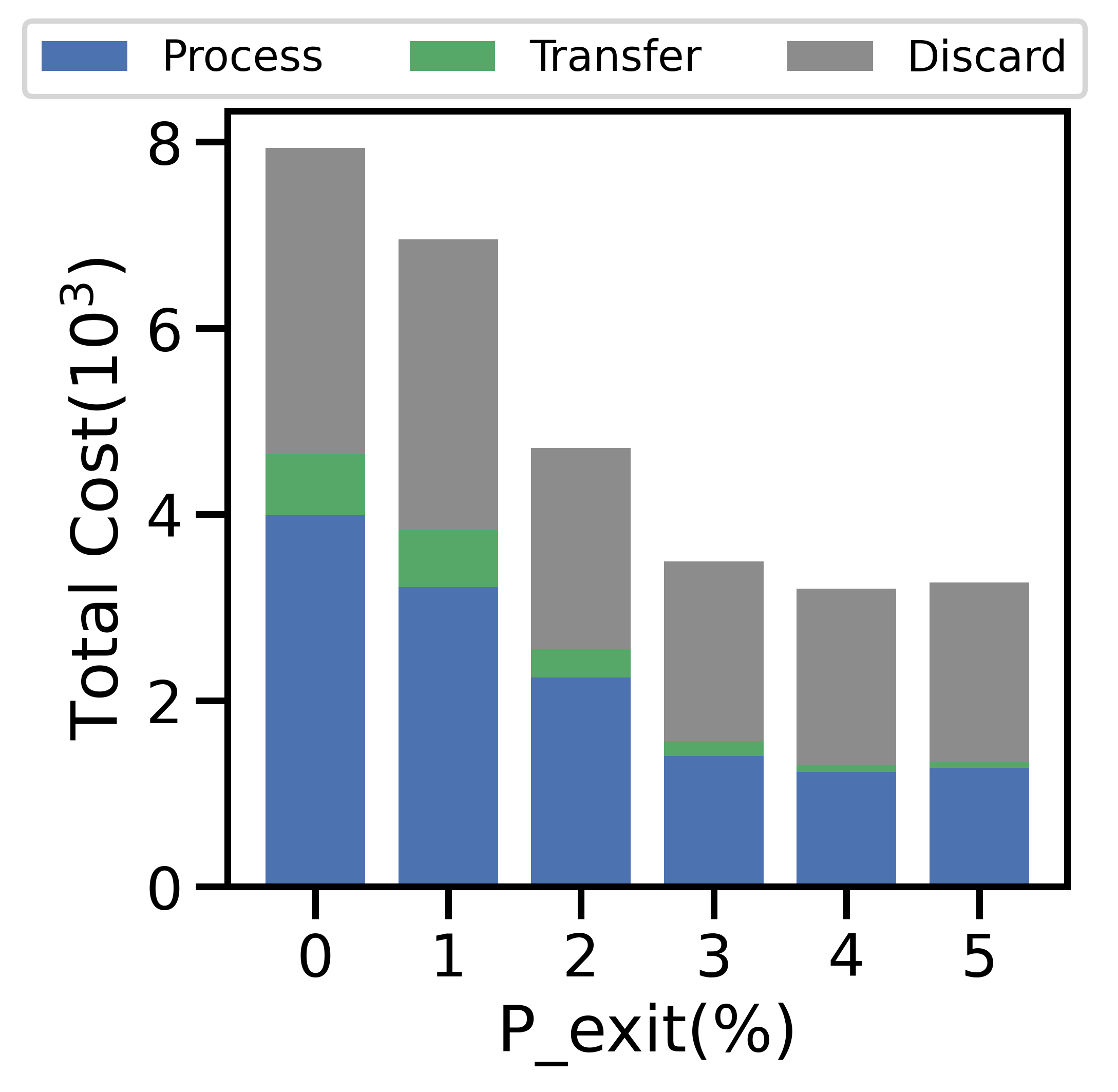}
\caption{Total cost breakdown}
\label{fig: varying_dynamic_exit_c}
\end{subfigure}
\begin{subfigure}[t]{0.198\textwidth} 
\centering
\includegraphics[height=1.3in,scale=0.9]{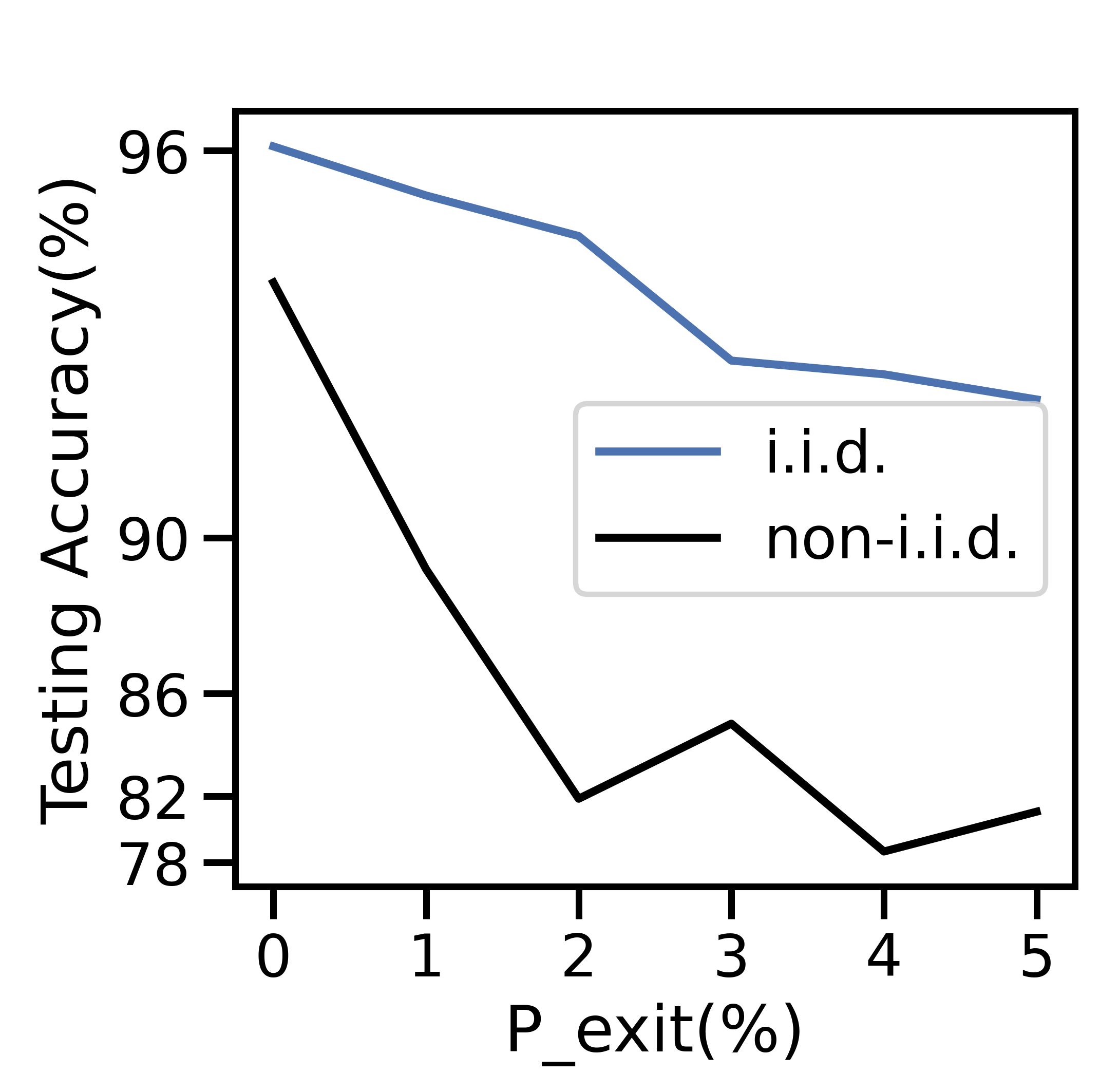} 
\caption{Testing accuracy}
\label{fig: varying_dynamic_exit_d}
\end{subfigure}
\caption{Impact of increases in the probability of node exits on data movement and costs with probability of node re-entry at 2\%. The shading in (a) and (c) indicates the range over time periods. While the i.i.d. case is more robust to dynamic network environments, both i.i.d. and non-i.i.d. cases indicate the same decreasing trends in testing accuracy as $p_{exit}$ grows.}
\label{fig: varying_dynamic_exit}
\vspace{-0.15in}
\end{figure*}

\begin{figure*}[t]
\centering
\captionsetup[subfigure]{oneside,margin={0.1cm,0.1cm}}
\begin{subfigure}[b]{0.175\textwidth}
\centering
\includegraphics[height=1.3in,scale=0.9]{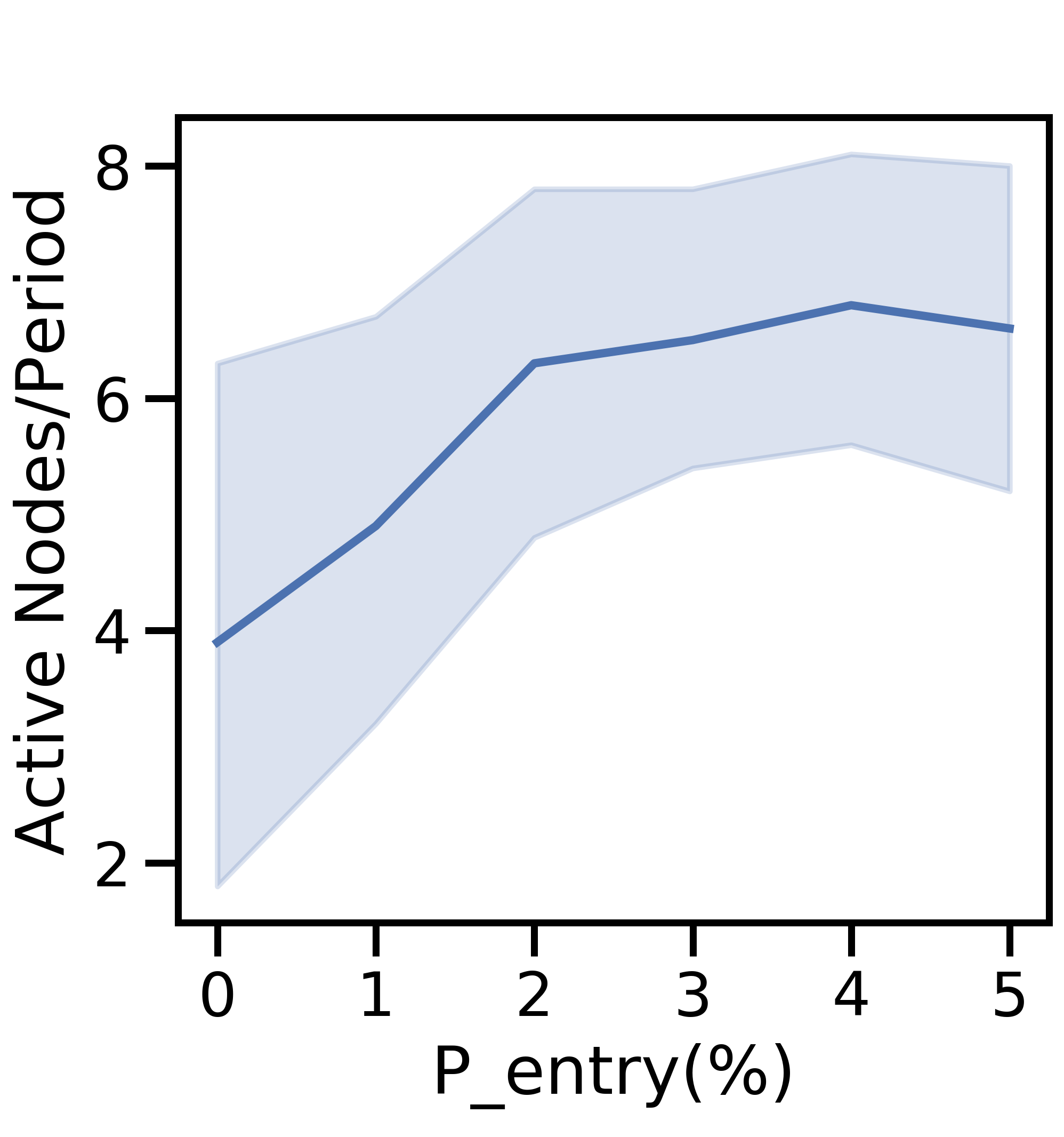}
\caption{Mean active nodes}
\label{fig: varying_dynamic_entry_e}
\end{subfigure}
\begin{subfigure}[b]{0.193\textwidth}
\centering
\includegraphics[height=1.3in,scale=0.9]{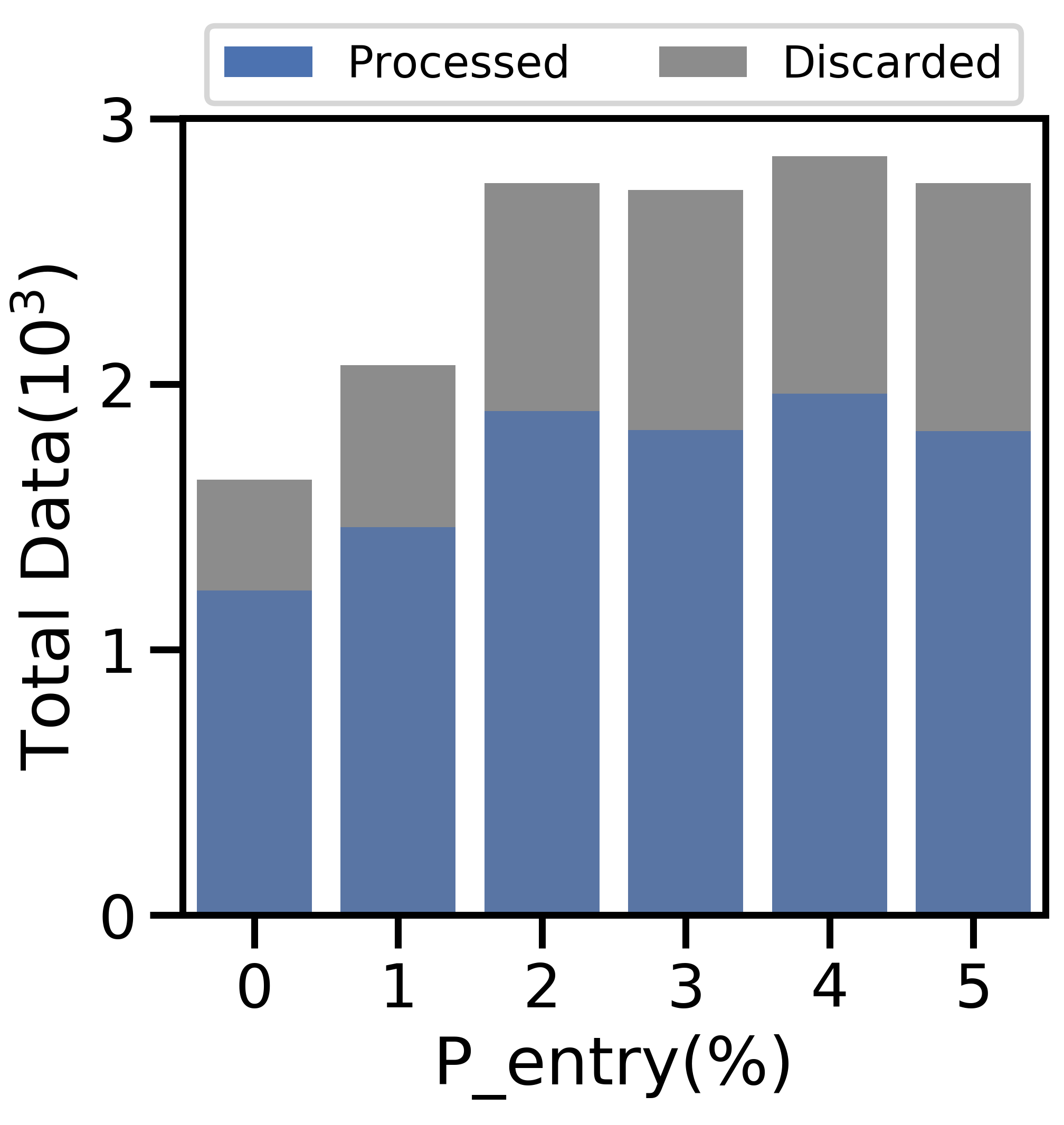}
\caption{Process vs. discard}
\label{fig: varying_dynamic_entry_a}
\end{subfigure} 
\begin{subfigure}[b]{0.2\textwidth} 
\centering
\includegraphics[height=1.3in,scale=0.9]{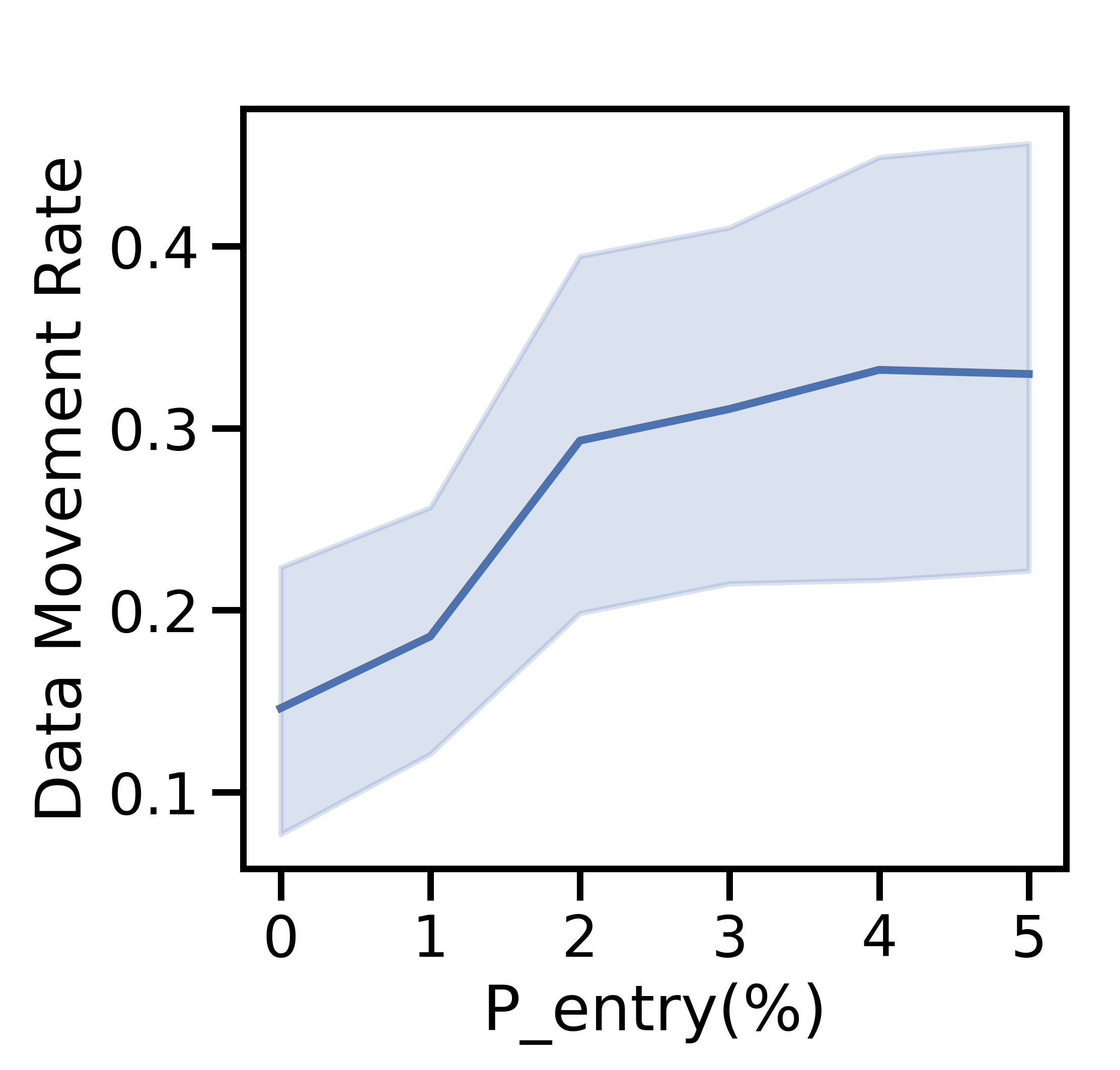}
\caption{Data movement rates}
\label{fig: varying_dynamic_entry_b}
\end{subfigure}
\begin{subfigure}[b]{0.2\textwidth}
\centering
\includegraphics[height=1.3in,scale=0.9]{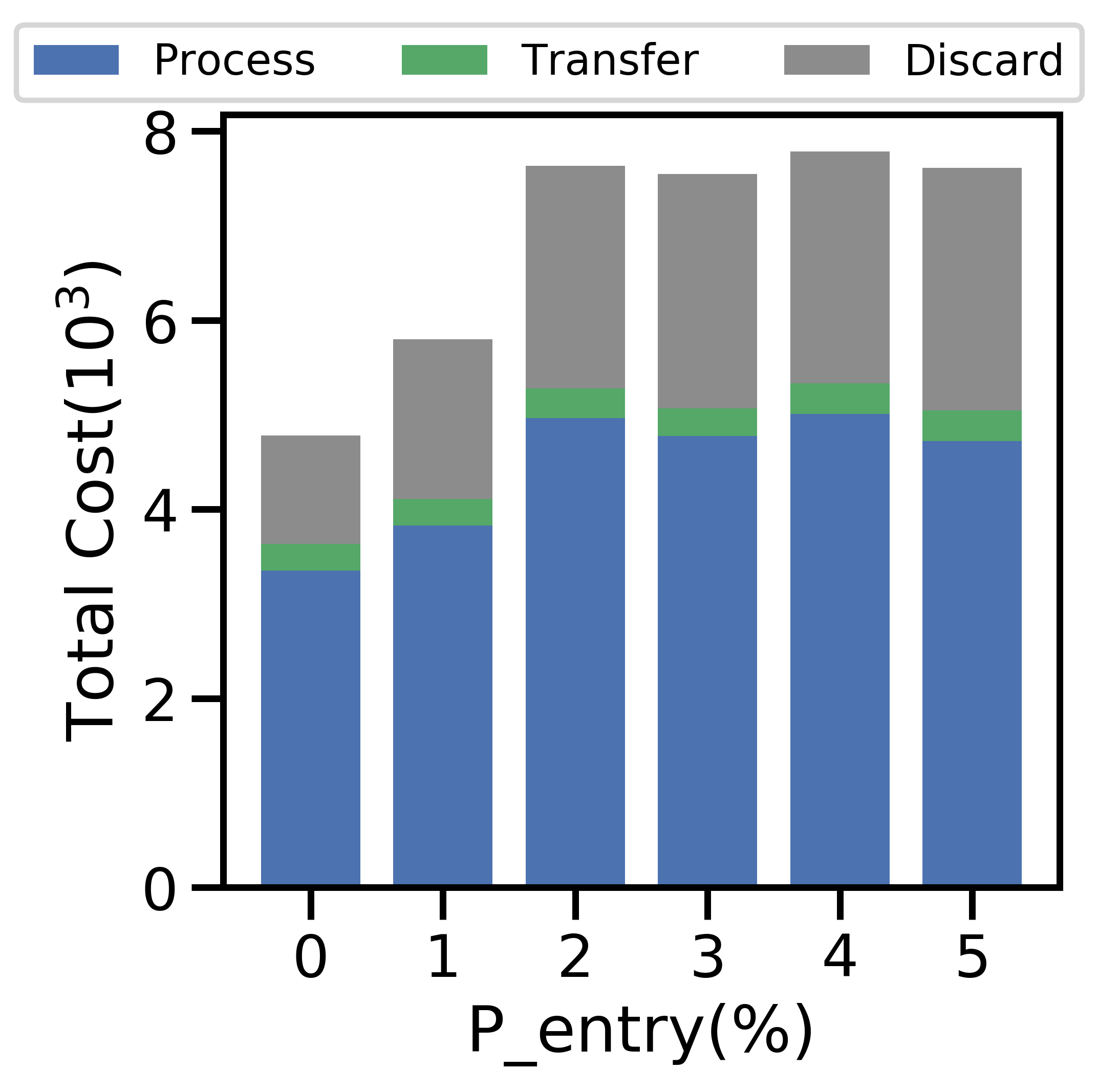}
\caption{Total cost breakdown}
\label{fig: varying_dynamic_entry_c}
\end{subfigure}
\begin{subfigure}[b]{0.195\textwidth}
\centering
\includegraphics[height=1.3in,scale=0.9]{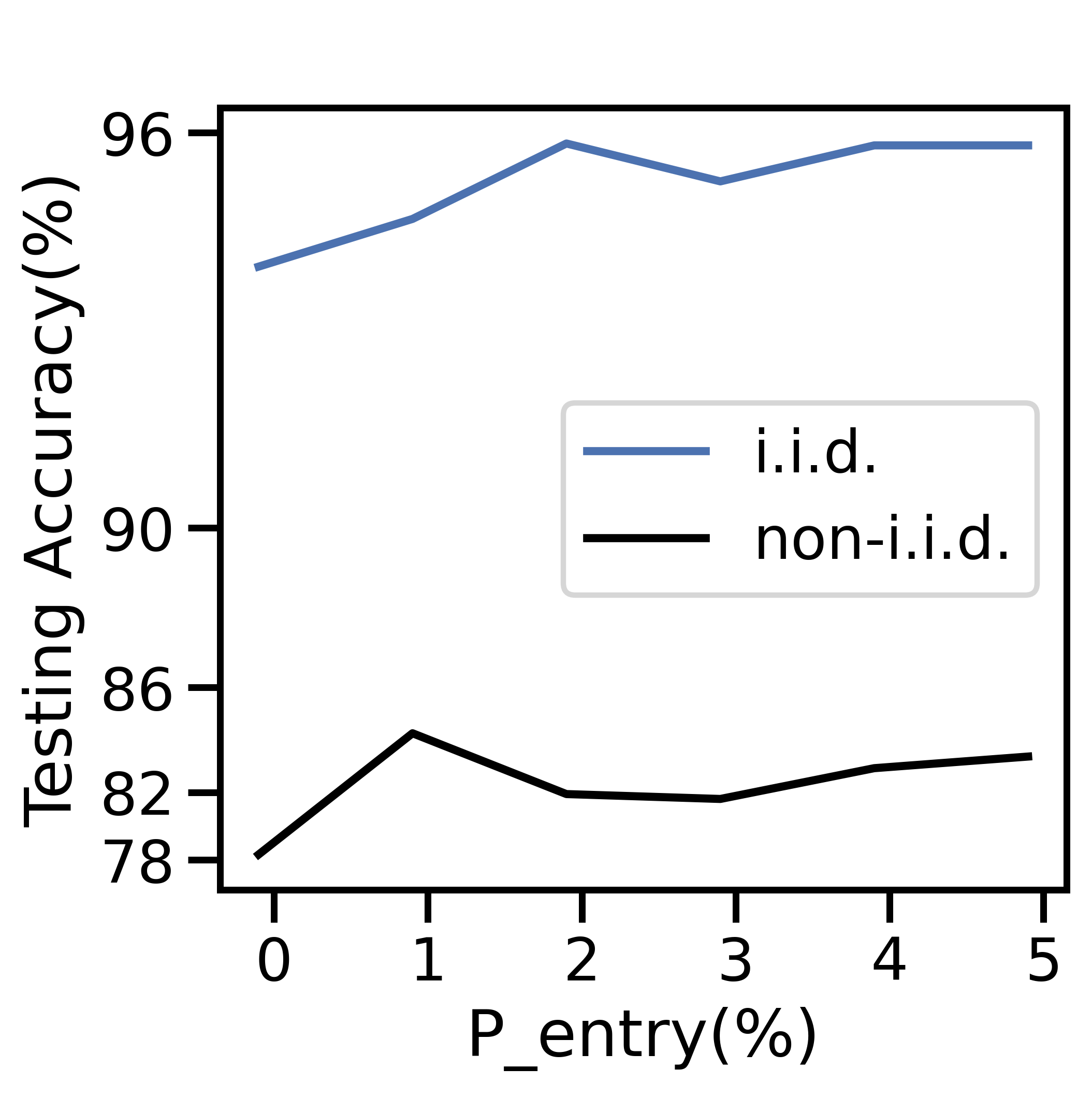}
\caption{Testing accuracy}
\label{fig: varying_dynamic_entry_d}
\end{subfigure}
\caption{Impact of increases in the probability of node re-entry in each time period, with the probability of node exits fixed at 2\%. The shading in (a) and (c) indicates the range over time periods. Both i.i.d. and non-i.i.d. cases improve in performance as $p_{entry}$ increases.}
\label{fig: varying_dynamic_entry}
\vspace{-0.2in}
\end{figure*}

\subsubsection{Varying $p_{exit}$} Figure \ref{fig: varying_dynamic_exit} varies $p_{exit}$ from 0-5\% and fixes $p_{entry} = 0.02$. Figure \ref{fig: varying_dynamic_exit}(a) shows the variation in average active nodes per period, and the remaining four subfigures display the aspects of network-aware learning as in Figs. 5-7.

Figure~\ref{fig: varying_dynamic_exit}(a) depicts a sharp decline in the number of active nodes per period as $p_{exit}$ grows. At $p_{exit} = 5\%$, the network averages $< 6$ active nodes/period, a 40\% decrease from network initialization. The total generated data and cost decrease sharply in Figures~\ref{fig: varying_dynamic_exit}(b) and \ref{fig: varying_dynamic_exit}(c), respectively: since the network has fewer active nodes, there is less data overall and therefore less cost. While the ratio of processed to discarded data in Figure~\ref{fig: varying_dynamic_exit}(b) skews towards processed data as $p_{exit}$ increases, the total cost in Figure~\ref{fig: varying_dynamic_exit}(d) skews toward discard costs: fewer active nodes implies fewer offloading opportunities, and network-aware learning discards more data. 
As a result, the average data movement rate drops from $0.55$ to $0.2$ in Figure \ref{fig: varying_dynamic_exit}(c). The i.i.d. testing accuracy declines by $\sim4$\% in Figure~\ref{fig: varying_dynamic_exit}(e), with a larger decline for non-i.i.d. data. 

\subsubsection{Varying $p_{entry}$} Figure~\ref{fig: varying_dynamic_entry} varies $p_{entry}$ from 0-5\% with $p_{exit} = 2\%$, and the average active nodes per period increases with probability of node entry until 4\% in Figure~\ref{fig: varying_dynamic_entry}(a). 
Both the total data generated (Figure~\ref{fig: varying_dynamic_entry}(b)) and average data movement rate (Figure~\ref{fig: varying_dynamic_entry}(c)) increase with more active nodes, but more data leads to larger processing, discard, and total costs in Figure~\ref{fig: varying_dynamic_entry}(d). 
As $p_{entry}$ increases, the network benefits from scale, i.e., more efficient offloading, which enables the network to train a more accurate ML model for the i.i.d. and non-i.i.d. cases in Figure~\ref{fig: varying_dynamic_entry}(e). 
However, the overall non-i.i.d. testing accuracy is unable to match that of the i.i.d. case, as node exits distort the distribution of local data across active devices further from the true global distribution. 

Figures~\ref{fig: varying_dynamic_exit} and \ref{fig: varying_dynamic_entry} exhibit a consistent pattern: the costs rapidly change and then plateau around when $p_{exit} \geq p_{entry}$ for Figure~\ref{fig: varying_dynamic_exit} and $p_{entry} \geq p_{exit}$ for Figure~\ref{fig: varying_dynamic_entry}. For the non-i.i.d. case, we notice that $p_{exit}$ has a stronger effect on the testing accuracy in Figure~\ref{fig: varying_dynamic_exit} than $p_{entry}$ does in Figure~\ref{fig: varying_dynamic_entry}; in the non-i.i.d. case, there is less overlap between devices' datasets, so each node exit affects the empirical training dataset more substantially and results in a stronger impact on the accuracy.

\section{Conclusion and Future Work}
\label{sec:conclusion}

In this paper, we developed a novel methodology to distribute ML training tasks over devices in a fog computing network while considering the compute-communication-accuracy tradeoffs inherent in fog scenarios. 
We derived new error bounds when devices transfer their local data processing to each other, and bounded the impact of these transfers on the cost and accuracy of the model training. Through experimentation with a popular machine learning task, we showed that our network-aware scheme significantly reduces the cost of model training while achieving comparable accuracy to the recently popularized federated learning algorithm for distributed training, and demonstrated the effects of network characteristics and device data distributions on its performance.

Our framework and analysis point to several possible extensions. First, while we do not observe significant heterogeneity in compute times on our wireless testbed, in general fog devices may experience compute straggling and failures, which might benefit from more sophisticated offloading mechanisms. Second, predicting devices' mobility patterns and network connectivity could likely further optimize the data offloading. Finally, learning device-specific models when local distributions are non-i.i.d. could introduce new performance tradeoffs between offloading and data processing.

\section*{Acknowledgment}
This work was partially supported by NSF CNS-1909306, by ARO grant W911NF1910036, and by NSWC Crane. We thank the anonymous reviewers for their valuable comments.

\balance
\bibliographystyle{IEEEtran}
{\small \bibliography{Reference}}

\begin{IEEEbiographynophoto}{Su Wang} is a second year PhD student in Electrical and Computer Engineering at Purdue University. He received his BS in Electrical Engineering from Purdue in 2018.
\end{IEEEbiographynophoto}
\vspace{-0.5cm}
\begin{IEEEbiographynophoto}{Yichen Ruan} is a Ph.D. candidate in Electrical and Computer Engineering at Carnegie Mellon University. He received his B.S. and M.S. degrees respectively from Tsinghua University and UC Berkeley.
\end{IEEEbiographynophoto}
\vspace{-0.5cm}
\begin{IEEEbiographynophoto}{Yuwei Tu} is an independent researcher conducting research on machine learning and network optimization. She graduated from the Center of Data Science at New York University with a master degree of data science.
\end{IEEEbiographynophoto}
\vspace{-0.5cm}
\begin{IEEEbiographynophoto}{Satyavrat Wagle} is a researcher in the Software Systems and Services Research Area (SSS-RA) at Tata Consultancy Services (TCS) Research. He holds an MS in Electrical Engineering from Carnegie Mellon University.
\end{IEEEbiographynophoto}
\vspace{-0.5cm}
\begin{IEEEbiographynophoto}{Christopher G. Brinton} (SM'20) is an Assistant Professor of Electrical and Computer Engineering at Purdue University. He received his PhD degree in Electrical Engineering from Princeton University in 2016. Since joining Purdue in 2019, he has won several awards including the \textit{Seed for Success Award} and the \textit{Ruth and Joel Spira Outstanding Teacher Award}.
\end{IEEEbiographynophoto}
\vspace{-0.5cm}
\begin{IEEEbiographynophoto}{Carlee Joe-Wong} (M'16) is the Robert E. Doherty Assistant Professor of Electrical and Computer Engineering at Carnegie Mellon University. She received her A.B., M.A., and Ph.D. degrees from Princeton University in 2011, 2013, and 2016, respectively. She has received several awards for her work, including the Army Young Investigator and NSF CAREER awards.
\end{IEEEbiographynophoto}

\newpage

% !TEX root = ../main.tex

\appendix

\section{Proofs of All Theorems}

\subsection{Proof of Theorem 1}%\ref{thm:error}}
\label{ref: proof_error}
\begin{IEEEproof}
To aid in analysis, we define $v_{k}(t) = v_{k}(t - 1) - \eta \nabla L(v_{k}(t - 1) | \mathcal{D})$ for $t \in \{(k-1)\tau, ..., k\tau\}$ as the centralized version of the gradient descent update synchronized with the weighted average $w(t)$ after every global aggregation $k$, i.e., $v_{k+1}(k \tau) \leftarrow w(k)$. With $\theta_k(t) = L(v_k(t)) - L(w^{\star})$, letting $K = \lfloor t / \tau \rfloor$, we can write %express $\phi_{K+1}(t) = \frac{1}{\theta_{K+1}(t)} - \frac{1}{\theta_{1}(0)}$ as
\begin{align}
\frac{1}{\theta_{K+1}(t)} &- \frac{1}{\theta_{1}(0)} = \Big( \frac{1}{\theta_{K+1}(t)} - \frac{1}{\theta_{K+1}(K\tau)} \Big) \nonumber \\
&+ \Big( \frac{1}{\theta_{K+1}(K\tau)}  - \frac{1}{\theta_{K}(K\tau)} \Big) + \Big( \frac{1}{\theta_{K}(K\tau)} - \frac{1}{\theta_{1}(0)} \Big) \nonumber \\
&\geq \Big( (t - K\tau) \omega \eta \big(1 - \frac{\beta \eta}{2} \big) \Big) + \Big( - \frac{\rho h(\tau)}{\epsilon^2} \Big) \nonumber \\
&\qquad + \Big (K \tau \omega \eta \big(1 - \frac{\beta \eta}{2} \big) - (K - 1) \frac{\rho h(\tau)}{\epsilon^2} \Big) \nonumber \\
&= t \omega \eta \Big( 1 - \frac{\beta \eta}{2} \Big) - K \frac{\rho h(\tau)}{\epsilon^2} \label{eq:thm1:b1}
\end{align}
where the three inequalities use the results $\frac{1}{\theta_k (k\tau)} - \frac{1}{\theta_k((k-1)\tau)} \geq \tau \omega \eta (1 - \frac{\beta \eta}{2} )$, $\frac{1}{\theta_{k+1}(k\tau)} - \frac{1}{\theta_k(k\tau)} \geq - \frac{\rho h(\tau)}{\epsilon^2}$, and $\frac{1}{\theta_K(T)} - \frac{1}{\theta_1(0)} \geq T \omega \eta ( 1 - \frac{\beta \eta}{2}) - (K - 1) \frac{\rho h(\tau)}{\epsilon^2}$ from Lemma 2 in \cite{wang2019adaptive}. Here, $\omega = \min_k \frac{1}{|| v_k( (k-1) \tau) - w^{\star}||^2}$ and $h(x) = \frac{\delta}{\beta} ((\eta \beta + 1)^x - 1) - \eta \delta x$ for $x \in \{0, 1, ...\}$. Additionally, if we assume $\theta_k(k\tau) = L(v_k(k\tau)) - L(w^{\star}) \geq \epsilon$, we can write
\begin{align}
&\frac{1}{L(w_i(t)) - L(w^{\star})} - \frac{1}{\theta_{K+1}(t)} \nonumber \\
&= \frac{\theta_{K+1}(t) - \big( L(w_i(t)) - L(w^{\star}) \big)}{\big(L(w_i(t)) - L(w^{\star}) \big) \theta_{K+1}(t)} \nonumber \\
&= \frac{L(v_{K+1}(t)) - L(w_i(t))}{\big( L(w_i(t)) - L(w^{\star}) \big) \theta_{K+1}(t)} \geq - \frac{\rho g_i(t - K\tau)}{\epsilon^2} \label{eq:thm1:b2}
\end{align}
where the inequality uses the result $||w_i(t) - v_k(t)|| \leq g_i(t - (k - 1)\tau)$ for any $k$ from Lemma 3 in \cite{wang2019adaptive}, and the $\rho$-Lipschitz assumption on $L_i(w)$ which extends to $L(w)$ by the triangle inequality, i.e., $|| L(x) - L(y) || \leq \rho ||x - y||$ for any $x, y$. Adding the results from (\ref{eq:thm1:b1}) and (\ref{eq:thm1:b2}) and noting $\theta_k(t) \geq 0$, we have
\begin{align}
& \frac{1}{L(w_i(t)) - L(w^{\star})} \geq \frac{1}{L(w_i(t)) - L(w^{\star})} - \frac{1}{\theta_1(0)} \nonumber \\
& \qquad \quad \geq t \omega \eta \big( 1 - \frac{\beta \eta}{2} \big) - \frac{\rho}{\epsilon^2} \big( Kh(\tau) + g_i(t - K\tau) \big) \nonumber
\end{align}
Taking the reciprocal, it follows that
\begin{align}
& L(w_i(t)) - L(w^{\star}) \leq \nonumber \\
& \frac{1}{t \omega \eta \big( 1 - \frac{\beta \eta}{2} \big) - \frac{\rho}{\epsilon^2} \big( Kh(\tau) + g_i(t - K\tau) \big)} = y(\epsilon)
\end{align}
for $y(\epsilon) > 0$. Now, let $\epsilon_0$ be the positive root of $y(\epsilon) = \epsilon$, which is easy to check exists. We can show that one of the following conditions must be true: (i) $\underset{k \leq K}{\min} \; L(v_k(k\tau)) - L(w^{\star}) \leq \epsilon_0$ or (ii) $L(w_i(t)) - L(w^{\star}) \leq \epsilon_0$. If we assume $L(v_k(k\tau))$ is non-increasing with $k$, then from (i) and (ii) we can write $L(w_i(t)) \leq L(w^{\star}) + \epsilon_0$ or $L(v_{K+1}(K\tau)) \leq L(w^{\star}) + \epsilon_0$. But we already know $||w_i(t) - v_k(t)|| \leq g_i(t - (k - 1)\tau)$ for any $k$, so with the $\rho$-Lipschitz assumption $L(w_i(t)) - L(v_{K+1}(t)) \leq \rho g_i(t - K\tau)$. If (ii) holds, then, $L(w_i(t)) \leq L(w^{\star}) + \epsilon_0 + \rho g_i(t - K\tau)$. Now comparing (i) and (ii), (ii) must always be true since $\rho, g_i \geq 0$.
\end{IEEEproof}

\subsection{Proof of Theorem 2}%\ref{thm:compute-exp}}
\label{ref: proof_compute}

\begin{IEEEproof}
Since we assume all costs and capacities are constant, we first note that the $G_i(t)$ can also be assumed to be constant. Thus, the processing of data at node $i$ can be modeled as a D/M/1 queue, with constant arrival rate $1/G_i(t)$ and exponential service time. The expected waiting time of such a queue is then equal to $\delta/\left(\mu(1 - \delta)\right)$, where $\delta$ is the smallest solution to $\delta = \exp\left(-\mu(1 - \delta)/G_i(t)\right)$. Upon showing that the expected waiting time is an increasing function of $\delta$ and $\delta$ an increasing function of $G_i(t)$, it follows that to ensure an expected waiting time no larger than $\sigma$, we should choose $G_i(t) \leq C$, where $C$ is the maximum arrival rate such that $\delta(C)/\left(\mu(1 - \delta(C))\right) = \sigma$.
\end{IEEEproof}

%\subsection{Proof of Theorem~\ref{thm:network-exp}}
%\label{ref: proof_network}
%
%\begin{IEEEproof}
%We prove the result by noting that the average total processing time is simply the sum of the average transmission time from device $j$ to $i$ and the average compute time at device $i$.
%To compute the average transmission time, we can model the process of sending data from each device $j$ to device $i$ by a M/M/1 queue; thus, the expected time for data to arrive at device $i$ from device $j$, i.e., the expected transmission time, is $1/(\mu_{ij} - \lambda_{ij})$. Since the departure rate from such a queue is another M/M/1 queue with the same arrival rate~\cite{bean1998output} and the union of multiple Poisson processes is another Poisson process, the computations at device $i$ can be viewed as a M/M/1 or M/D/1 queue with arrival rate $\sum_{j\in \cup \left(i, J(i)\right)} \lambda_{ij}$. Then following logic similar to that in Theorem~\ref{thm:compute-exp}, we can model the data processing at each node as an M/D/1 or M/M/1 queue when processing times at the node are deterministic or exponential, respectively. The result follows from the known system wait time distributions of both types of queues.
%\end{IEEEproof}

\subsection{Proof of Theorem 4}%~\ref{thm:nonlinear}}
\label{ref: proof_nonlinear}

\begin{IEEEproof}
In the hierarchical scenario described in the statement of the theorem, the cost objective (\ref{eq:dataOpt}) can be rewritten as
\begin{eqnarray*}
\sum_i (1 - r_i - s_i ) D_i c_i + \sum_i s_i D_i (c_{n+1} + c_t) \\ + \sum_i \frac{\gamma}{\sqrt{(1 - r_i - s_i)D}} + \frac{\gamma}{\sqrt{\sum_i s_i D_i}}.
\end{eqnarray*}
Taking the partial derivative of the cost objective with respect to $r_i$ and setting to 0 gives:
\begin{equation*}
-D_i c_i + \frac{\gamma (1/2) D_i}{ ((1-r_i-s_i) D_i)^{3/2} } = 0
\end{equation*}
Rearranging gives $2c_i = \frac{\gamma}{((1-r_i-s_i)D_i)^{3/2}}$, which yields:
\begin{equation*}
r_i^* = 1 - \frac{{(\gamma / 2c_i)}^{2/3}}{D_i} - s_i. 
\end{equation*}
Using the expression for $r_i^*$, the objective function becomes:
\begin{eqnarray*}
\sum_i (1 - r_i^* - s_i ) D_i c_i + \sum_i s_i D_i (c_{n+1} + c_t) \\ + \sum_i \frac{\gamma}{\sqrt{(1 - r_i^* - s_i)D}} + \frac{\gamma_{n+1}}{\sqrt{\sum_i s_i D_i}}.
\end{eqnarray*}
Taking the partial derivative with respect to $s_i$ and setting to $0$ gives:
\begin{eqnarray*}
D_i c_i (-\frac{d r_i^*}{ds_i} - 1) + D_i(c_{n+1}+c_t) \\ - \frac{\gamma D_i (-\frac{dr_i^*}{ds_i}-1)}{2((1-r_i-s_i)D_i)^{3/2}} - \frac{\gamma D_i}{2(\sum_j s_j D_j)^{3/2}} = 0.
\end{eqnarray*}
Since $r_i + s_i = 1$ at the optimal point, $\frac{dr_i^*}{ds_i} = -1$ and we obtain $D_i(c_{n+1}+c_t) - \frac{\gamma D_i}{2(\sum_j s_j D_j)^{3/2}} = 0$ for each $i$. We can then enforce $s_i = s_j$ for all $i,j$ and rearrange for $s_i$ to obtain: 
\begin{equation*}
s_i^* = \frac{1}{\sum_j D_j} \left(\frac{\gamma}{2(c_{n+1}+c_t)}\right)^{2/3}
\end{equation*}
Finally, note that a large $D_i$ forces $r_i, s_i \in [0, 1]$, which gives the result.
\end{IEEEproof}

\subsection{Proof of Theorem 5} %\ref{thm:cost_uniform}}
\label{ref: proof_cost}

\begin{IEEEproof}
From Theorem 3, we can write the expected cost savings as %~\ref{thm:unconstrained}
\begin{equation}
\resizebox{0.5 \textwidth}{!}
{
$ \mathbb{E}\left[\max \left\{0, c_i - \min_j c_j\right\}\right] = \int_0^C \int_0^{c_i} \frac{k(c_i - y)}{C^2}\left(\frac{C - y}{C}\right)^{k - 1}\;dy\;dc_i $
}
\end{equation}
where we take $y = \min_j c_j$ and use the fact that the minimum of $k$ i.i.d. uniform random variables on $[0,C]$ has the probability distribution $f(y) = \frac{k}{C}\left(\frac{C - y}{C}\right)^{k - 1}$. Integrating and simplifying then yields the desired result.
\begin{align*}
&\int_0^C \int_0^{c_i} \frac{k(c_i - y)}{C^2}\left(\frac{C - y}{C}\right)^{k - 1}\;dy\;dc_i \\
&= \frac{k}{C^{k + 1}} \int_0^C \int_0^{c_i} c_i\left(C - y\right)^{k - 1} - y\left(C - y\right)^{k - 1} \;dy\;dc_i \\
&= \frac{k}{C^{k + 1}} \int_0^C \Bigg(\frac{c_i}{k}\left(C^k - (C - c_i)^k\right) \\
&+ \int_0^{c_i}\sum_{l = 0}^{k - 1} \binom{k - 1}{l} (-y)^{l + 1} C^{k - 1 - l}\;dy\Bigg)\;dc_i \\
&= \frac{k}{C^{k + 1}} \Bigg(\frac{C^{k + 2}}{2k} + \frac{1}{k}\int_0^C \Bigg(\sum_{l = 0}^k \binom{k}{l} (-c_i)^{l + 1} C^{k - l} \\
&+ \sum_{l = 0}^{k - 1} \binom{k - 1}{l} \frac{{c_i}^{l + 2}C^{k - 1 - l}(-1)^{l + 1}}{l + 2}\Bigg)\;dc_i\Bigg) \\
&= \frac{C}{2} + \sum_{l = 0}^k \binom{k}{l} (-1)^{l + 1} C^{-l - 1}\frac{C^{l + 2}}{l + 2} \\
&+ \sum_{l = 0}^{k - 1} \binom{k - 1}{l} \frac{k C^{l + 3}C^{k - 1 - l}(-1)^{l + 1}}{(l + 2)(l + 3)C^{k + 1}} \\
&= \frac{C}{2}  + \sum_{l = 0}^k \binom{k}{l} \frac{(-1)^{l + 1} C}{l + 2} + \sum_{l = 0}^{k - 1} \binom{k - 1}{l} \frac{k C(-1)^{l + 1}}{(l + 2)(l + 3)} \\
&= \frac{C}{2} - \frac{C(-1)^k}{k + 2} - \sum_{l = 0}^{k - 1} \binom{k}{l} \frac{C(-1)^l(k + 3)}{(l + 2)(l + 3)}.
\end{align*}

\end{IEEEproof}

\end{document}